\documentclass[seceq]{ptptex}

\usepackage{graphicx}
\usepackage{wrapft}

\notypesetlogo                       
\preprintnumber{
hep/ph-yymmnnn\\STUPP-07-189}

\title{
  An Illustration and Analysis of the Degeneracy in the Search
  for the Leptonic $CP$-Violating Phase and the Neutrino Mass Hierarchy
}

\author{
  Masafumi \textsc{Koike}\footnote{%
    E-mail: koike@krishna.th.phy.saitama-u.ac.jp}
  and
  Masako \textsc{Saito}\footnote{%
    E-mail: msaito@krishna.th.phy.saitama-u.ac.jp}
}

\inst{
  Physics Department, Saitama University,
  Saitama 338-8570, Japan
}

\abst{
  Determination of the value of the leptonic $CP$-violating phase
  $\delta$ and the neutrino mass hierarchy $\mathrm{sgn} \, \delta
  m^{2}_{31}$ through long baseline neutrino oscillation experiments
  is systematically analyzed.
  We note that the two oscillation spectra are difficult to
  discriminate and lead to the degeneracy when they are peaked at the
  same energy and have the same peak probability.
  The condition of peak-matching is therefore introduced as a
  criterion for the presence of degeneracy.
  The matching of peaks is visualized as an intersection of
  trajectories traced by the peak of an oscillation spectrum while the
  value of $\delta$ is varied from $0$ to $2\pi$.
  We numerically calculate a pair of trajectories for a pair of
  hierarchies and examine the degeneracy, especially that concerning
  the hierarchy.
  We formulate the trajectory in terms of analytic expressions and
  evaluate the critical length, which is shown to be proportional to
  $1/\sin \theta_{13}$.
  In view of our analysis, we discuss future prospects to solve the
  hierarchy degeneracy with regard to the following four approaches:
  elongating the baseline length sufficiently, using both neutrinos
  and anti-neutrinos, combining experiments with different baseline
  lengths, and observing two or more oscillation peaks.
}

\begin{document}

\maketitle


\section{Introduction}
Experiments suggest the existence of oscillation among the different
flavors of neutrinos, providing rich information on the flavor
structure of the lepton sector.%
\cite{bib:solar,bib:atmospheric,bib:terrestrial,Apollonio:2002gd} \
Despite this progress, however, present knowledge of neutrinos is
still incomplete.
In the notation of Ref.~\citen{Yao:2006px}, the values of the mass
parameters and the mixing parameters yet to be determined include one
of the mixing angles $\theta_{13}$, the sign of $\delta m^{2}_{31}$,
and the $CP$-violating phase $\delta$.
The value of $\theta_{13}$ is confined only by an upper bound as
$\sin^{2} 2\theta_{13} < 0.19$,\cite{Yao:2006px} while the sign of
$\delta m^{2}_{31}$ and the value of $\delta$ are completely unknown
to date.
$CP$ violation manifests itself only in the flavor-changing appearance
channel such as $\nu_{\mu} \to \nu_{\mathrm{e}}$, which in turn is
suppressed by a small factor of $\sin^{2} 2\theta_{13}$.
For this reason, we expect a two-staged program in pursuing the
unknown properties of neutrinos.
The first stage is the search for $\theta_{13}$.
It is anticipated that its upper bound will be improved down to
$\sin^{2} 2\theta_{13} \lesssim O(10^{-2})$ by planned experiments
using nuclear reactors \cite{bib:Nuclear-Planned} and accelerators.%
\cite{bib:Accelerator-Planned} \ 
The second stage is the search for the sign of $\delta m^{2}_{31}$ and
for $CP$ violation.
The next generations of long baseline neutrino oscillation experiments
offer promising opportunities for such searches,%
\cite{bib:Accelerator-Planned}
and other probes of the sign of $\delta m^{2}_{31}$ are also discussed
in the literature.\cite{bib:hierarchy-search} \

Our focus in this paper is on the second stage based on the optimistic
expectation that the next generation of reactor neutrino experiments
will find evidence of non-vanishing $\theta_{13}$ and that its value
will turn out to be large enough for the $CP$-violation search.
We consider the search for the $CP$-violating phase and the sign of
$\delta m^{2}_{31}$ through long baseline neutrino oscillation
experiments using a conventional beam of muon neutrinos.
%
%
A single long-baseline experiment, however, does not necessarily lead
to the determination of a unique set of values of oscillation
parameters.
One obstacle is the ambiguity of parameter values due to the
experimental errors.\cite{bib:ambiguity,Koike:2005dk} \ 
Another is the parameter degeneracy, which arises when the experimental
results can be attributed to two or more sets of parameter values.
Three types of the degeneracy are customarily referred to:
the intrinsic degeneracy, on $(\theta_{13}, \delta_{\mathrm{CP}})$;%
\cite{Burguet-Castell:2001ez}
the hierarchy degeneracy, on $\delta m^{2}_{31} \gtrless 0$;%
\cite{bib:biprob-plot}
and the octant degeneracy, on $\theta_{23} \gtrless \pi/4$.%
\cite{Fogli:1996pv} \ 
Although the parameter degeneracy should be avoided,%
\cite{%
Burguet-Castell:2001ez,%
bib:biprob-plot,%
Fogli:1996pv,%
bib:degeneracy,%
bib:two-baselines-combo,%
Barger:2001yr}
its presence is difficult to predict comprehensively
due to the complicated dependence of experimental results on many
parameters and on various setups of experiments.
%
%
%
%
The plot introduced in Ref.~\citen{bib:biprob-plot} gives an overview
of the presence of degeneracy and has been found versatile for its
analysis.
It presents the two oscillation probabilities or event rates of two
channels in a two-dimensional space, enabling to show separately the
$CP$-violating effect and the matter effect.

We introduce another aspect of the analysis of the energy spectrum of
the appearance probability.
Giving an intuitive illustration of the determination of the
parameters from this spectrum, we offer a view of ours on the
emergence of degeneracy and its resolution.
The pivot of our study is the peak of the oscillation probability,
particularly its position, or the energy and probability at the peak.
We note that two oscillation spectra whose peak positions coincide are
difficult to distinguish and likely to cause degeneracy.
We trace the peak position varying the values of $\delta$ and
$\mathrm{sgn} \, \delta m^{2}_{31}$ to show how their search is
entangled and the hierarchy degeneracy results.
We vary the baseline length as well and describe from our point of
view how the degeneracy disappears when the baseline becomes
sufficiently long.
We organize these analyses by deriving analytic expressions of the
oscillation probability at the peaks and offer an overview concerning
the presence of degeneracy and possible ways to avoid it.

The outline of this paper is as follows.
In \S\ref{sec:Peak-matching-cond}, we introduce the peak-matching
condition as a criterion for the presence of degeneracy and facilitate
its visualization by drawing closed trajectories of the oscillation
peak.
In \S\ref{sec:Peak-loops-in-action}, we develop an understanding of
the presence and absence of the degeneracy of parameters through the
loops of the peak.
In \S\ref{sec:Analytic-expressions}, we derive an analytic
expression of the $\nu_{\mu} \to \nu_{\mathrm{e}}$ appearance
probability at the oscillation peaks to elucidate its dependence on
the mass parameters and the mixing parameters, and show how the loops
are distorted due to the change of the baseline length.
In \S\ref{sec:Resolving-hierarchy-degeneracy}, we apply the peak loops
to the evaluation of four methods to uniquely determine the value of
the $CP$-violating phase and the mass hierarchy.
Section \ref{sec:Conclusion-and-discussions} presents the conclusion
and discussion.

\section{Peak-matching condition and the peak loop}
\label{sec:Peak-matching-cond}
We assume that there are three neutrino generations and adopt
the definitions given in Ref.~\citen{Yao:2006px} of the quadratic mass
differences $\delta m^{2}_{ij}$ ($\{i, j\} \subset \{1, 2, 3\}$), the
mixing angles $\theta_{ij}$, and the $CP$-violating phase $\delta$.
All experimental results obtained to this time can be accounted for by
neutrino oscillation by taking
\begin{math}
  \delta m^{2}_{21}
  \simeq (8.0^{+0.4}_{-0.3} ) \times 10^{-5} \, \mathrm{eV}^{2},
\end{math}
\begin{math}
  |\delta m^{2}_{31}|
  \simeq (1.9 \,\textrm{--}\, 3.0) \times 10^{-3} \, \mathrm{eV}^{2},
\end{math}
\begin{math}
  \sin^{2} 2\theta_{12} = 0.86^{+0.03}_{-0.04},
\end{math}
\begin{math}
  \sin^{2} 2\theta_{23} > 0.92,
\end{math}
and
\begin{math}
  \sin^{2} 2\theta_{13} < 0.19,
\end{math}
\cite{Yao:2006px} with the exception of the LSND
experiment,\cite{Aguilar:2001ty} whose results were not confirmed by
either the KARMEN experiment\cite{Armbruster:2002mp} or the MiniBooNE
experiment.\cite{Aguilar-Arevalo:2007it} \
We do not take account of the ambiguities of these parameters in this
paper for the sake of a clear presentation of our idea.
The influence of these ambiguities will be discussed in
\S\ref{sec:Conclusion-and-discussions}.
\begin{wrapfigure}[28]{r}{\halftext}
    \centerline{
      \includegraphics[width=65mm]{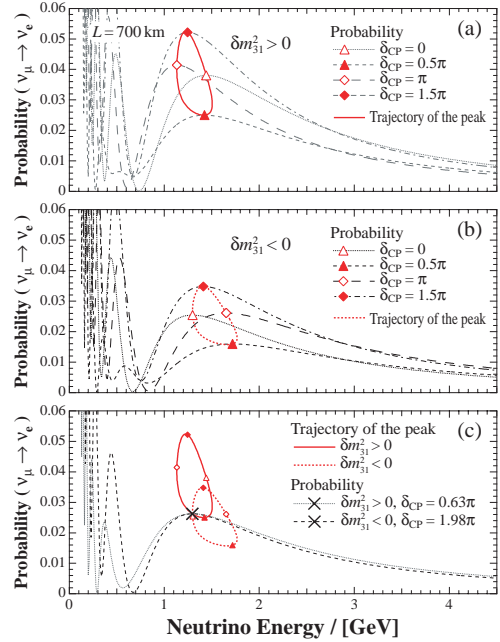}
    }
    \caption{
      The $\nu_{\mu} \to \nu_{\mathrm{e}}$ appearance 
      probabilities and
      the trajectories of their first peaks for a baseline length of
      $700 \, \mathrm{km}$.
      The parameter values in Eq.~(\ref{eq:example-params}) are
      adopted.
      Figure (a) is for the normal hierarchy and
      Fig.(b) is for the inverted.
      Each includes the probability
      spectra for $\delta = 0, \pi/2, \pi$ and $3\pi/2$.
      Figure (c) overlays the trajectories of the top two, along
      with two oscillation spectra peaked at the intersection
      of these trajectories marked by the ``$\times$'' symbol.
      %
      %
      %
    }
    \label{fig:appearance-probs}
\end{wrapfigure}

We consider the search for the $CP$-violating phase $\delta$ and the
mass hierarchy $\mathrm{sgn} \, \delta m^{2}_{31}$ by investigating
the $\nu_{\mu} \to \nu_{\mathrm{e}}$ appearance probability in long
baseline neutrino oscillation experiments.
We do not treat the disappearance channel, which is effective in
practical analyses to restrict parameter values such as the absolute
value of $\delta m^{2}_{31}$.
Let us first illustrate with an example that the $\nu_{\mu} \to
\nu_{\mathrm{e}}$ appearance probability enables to search the value
of $\delta$ and the sign of $\delta m^{2}_{31}$.
In Fig.~\ref{fig:appearance-probs}, we display the $\nu_{\mu} \to
\nu_{\textrm{e}}$ appearance probabilities in a case of a baseline
length of $L = 700 \, \mathrm{km}$.
There, the value of $\delta$ and the sign of $\delta m^{2}_{31}$ are
varied while other parameters are fixed to a set of example values: %
\begin{subequations}
\begin{align}
  \delta m^{2}_{21}
  & = 8.2 \times 10^{-5} \, \mathrm{eV}^{2} \, , \;
  \label{eq:example-params-dmm21}
  \\
  \bigl\lvert \delta m^{2}_{31} \bigr\rvert
  & = 2.5 \times 10^{-3} \, \mathrm{eV}^{2} \, , \;
  \label{eq:example-params-abs-dmm31}
  \\
  \sin^{2} 2\theta_{12} & = 0.84 \, , \;
  \label{eq:example-params-sin2-2theta12}
  \\
  \sin^{2} 2\theta_{23} & = 1.0 \, , \;
  \label{eq:example-params-sin2-2theta23}
  \\
  \sin^{2} 2\theta_{13} & = 0.06 \, .
  \label{eq:example-params-sin2-2theta13}
\end{align}
The matter density $\rho$ on the baseline is assumed to be constant
and fixed to
\begin{equation}
  \rho = 2.6 \, \mathrm{g/cm^{3}},
  \label{eq:example-params-rho}
\end{equation}
\label{eq:example-params}
\end{subequations}
which is related to the electron number density $n_{\textrm{e}}$ as
$n_{\textrm{e}} = N_{\textrm{A}} Y_{\textrm{e}} \rho$ with the
Avogadro constant $N_{\textrm{A}}$ and the proton-to-nucleon ratio
$Y_{\textrm{e}}$ on the baseline.
For convenience, we use this set of values and assume
$Y_{\textrm{e}} = 0.5$ for the numerical calculations in this paper
unless otherwise noted.
Figure \ref{fig:appearance-probs} plots the energy spectra of the
appearance probability (a) for $\delta m^{2}_{31} = +2.5 \times
10^{-3} \, \mathrm{eV^{2}} > 0$ (normal hierarchy) and (b) for $\delta
m^{2}_{31} = -2.5 \times 10^{-3} \, \mathrm{eV^{2}} < 0$ (inverted
hierarchy).
Figure \ref{fig:appearance-probs}(c) displays two spectra for the set
of parameter values given in it.
We can clearly see the dependence of the spectrum upon $\delta$ and
$\mathrm{sgn} \, \delta m^{2}_{31}$ in (a) and (b), and thus we can
search for these values through $\nu_{\mu} \to \nu_{\textrm{e}}$
appearance experiments.

This dependence, however, does not guarantee that we can uniquely
determine these values from a single experiment.
Experimental results in some cases are consistent with both normal and
inverted hierarchies, each with an appropriate choice of the value of
$\delta$.
For example, an experiment providing the oscillation spectrum of the
dashed curve in Fig.~\ref{fig:appearance-probs}(c) would reject neither
$(\mathrm{sgn} \, \delta m^{2}_{31}, \delta) = (+, 0.63 \pi)$ nor
$(\mathrm{sgn} \, \delta m^{2}_{31}, \delta) = (-, 1.98 \pi)$ and
leads to the hierarchy degeneracy, if the low-energy neutrinos below
about $1 \, \mathrm{GeV}$ are unobservable.
The degeneracy introduces unfavorable complication into the analysis
and should be avoided.

We now direct our attention to the peak of the appearance probability
spectrum since it gives insight into the presence of hierarchy
degeneracy.
An appearance probability has a series of peaks
\begin{math}
  ( E_{\textrm{peak}, n}, P_{\textrm{peak}, n} )
\end{math}
($n = 0, 1, 2, \cdots$) with $E_{\textrm{peak}, 0} > E_{\textrm{peak},
  1} > E_{\textrm{peak}, 2} > \cdots$.
Each peak traces out a closed trajectory as we vary the value of
$\delta$ from $0$ to $2 \pi$, keeping the other parameters fixed.
The loops traced by the first peak ($n = 0$) are presented in
Fig.~\ref{fig:appearance-probs}(a) for $\delta m^{2}_{31} > 0$ and (b)
for $\delta m^{2}_{31} < 0$.
The peak positions for $\delta = 0$, $\pi/2$, $\pi$, and $3\pi/2$ are
indicated on the loops by open triangles, solid triangles, open
diamonds, and solid diamonds, respectively.
We observe that the peak position for the normal (inverted) hierarchy
moves clockwise (counterclockwise) on the loop as the value of
$\delta$ increases.

In Fig.~\ref{fig:appearance-probs}(c), the peak loops for $\delta
m^{2}_{31} \gtrless 0$ are copied from
Figs.~\ref{fig:appearance-probs}(a) and (b) with the symbol
``$\times$'' marking one of their intersections.
This intersection corresponds to the two sets of parameter values
given in the figure.
With these values, the shown oscillation probabilities are both peaked at
the symbol as should be the case.
We observe the similarity of the two for $E \gtrsim 1 \, \mathrm{GeV}$
despite their difference for lower energies.
Given the typical visible energy of neutrinos $E > (0.5
\,\textrm{--}\, 1.0) \, \mathrm{GeV}$, we expect that their similarity
will make it difficult to distinguish the two sets of parameter values
by experiments and thus lead to degeneracy.

We arrive here at an intuitive understanding of the presence of
degeneracy.
Assume that the visible energy range of an experiment covers a peak
$(E_{\textrm{peak}, n}, P_{\textrm{peak}, n})$ of the appearance
probability. %
(In \S\ref{sec:Resolving-hierarchy-degeneracy} we discuss cases in which
two or more peaks are visible, which occurs when a sufficiently long
baseline, typically $L \gtrsim 1000 \, \mathrm{km}$, and a wide-band
neutrino beam are used.)
The values of $E_{\textrm{peak}, n}$ and $P_{\textrm{peak}, n}$ depend
on the oscillation parameters such as $\delta m^{2}_{ij}$,
$\theta_{ij}$, and $\delta$, which we collectively denote by $\{
\vartheta_{i} \}$.
The observation made in the previous paragraph signifies that the two
parameter sets $\{ \vartheta_{i} \}$ and $\{ \vartheta'_{i} \}$ will
be degenerate when the peak-matching condition
\begin{subequations}
\begin{align}
  E_{\textrm{peak}, n} \bigl( \{ \vartheta_{i} \} \bigr)
  & =  E_{\textrm{peak}, n} \bigl( \{ \vartheta'_{i} \} \bigr) \, ,
  \label{eq:epeak-matching-cond}
  \\
  P_{\textrm{peak}, n} \bigl( \{ \vartheta_{i} \} \bigr)
  & =  P_{\textrm{peak}, n} \bigl( \{ \vartheta'_{i} \} \bigr)
  \label{eq:ppeak-matching-cond}
\end{align}
\label{eq:peak-matching-cond}
\end{subequations}
is satisfied.
This condition was first tested in Ref.~\citen{Koike:2005dk} in the
analysis of an example study, although not extensively.
In particular, the hierarchy degeneracy arises when the values of
$\delta m^{2}_{31}$ in $\{ \vartheta_{i} \}$ and in $\{ \vartheta'_{i}
\}$ have opposite signs.
\begin{figure}
\parbox{\halftext}{
 \centerline{
   \includegraphics[width=65mm]{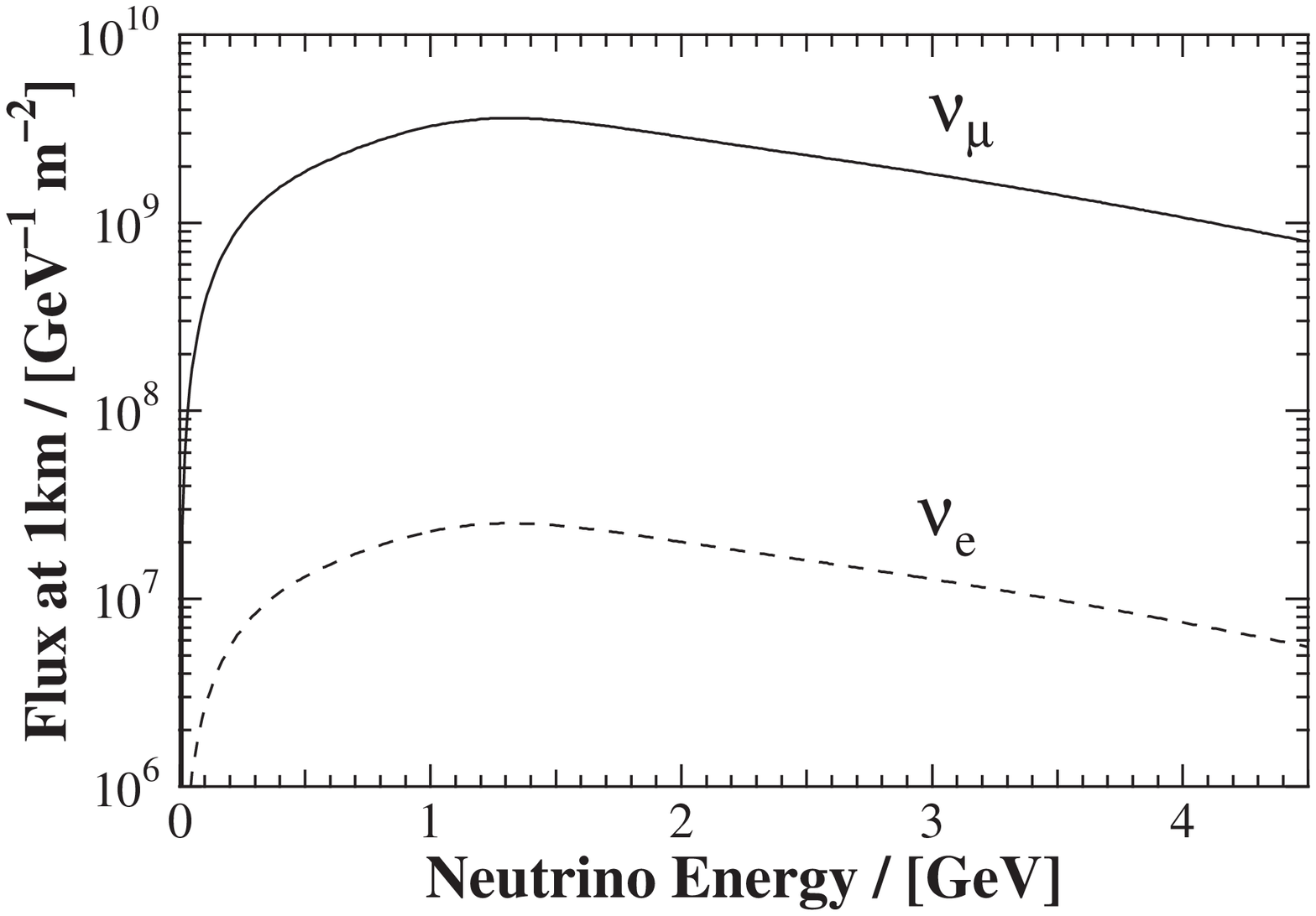}
 }
  \caption{ %
    The neutrino fluxes used in the example analysis.
    The flux of muon neutrinos (solid curve) is the spectrum of the
    wide-band neutrino beam expected from the upgraded Alternate
    Gradient Synchrotron at Brookhaven National
    Laboratory\cite{Diwan:2003bp}.
    The flux of electron neutrinos (dotted curve) is assumed to be 0.7\%
    of that of muon neutrinos with the same energy dependence.   
  }
  \label{fig:nu-flux}
}
\hfill
\parbox{\halftext}{%
  \centerline{
    \includegraphics[width=65mm]{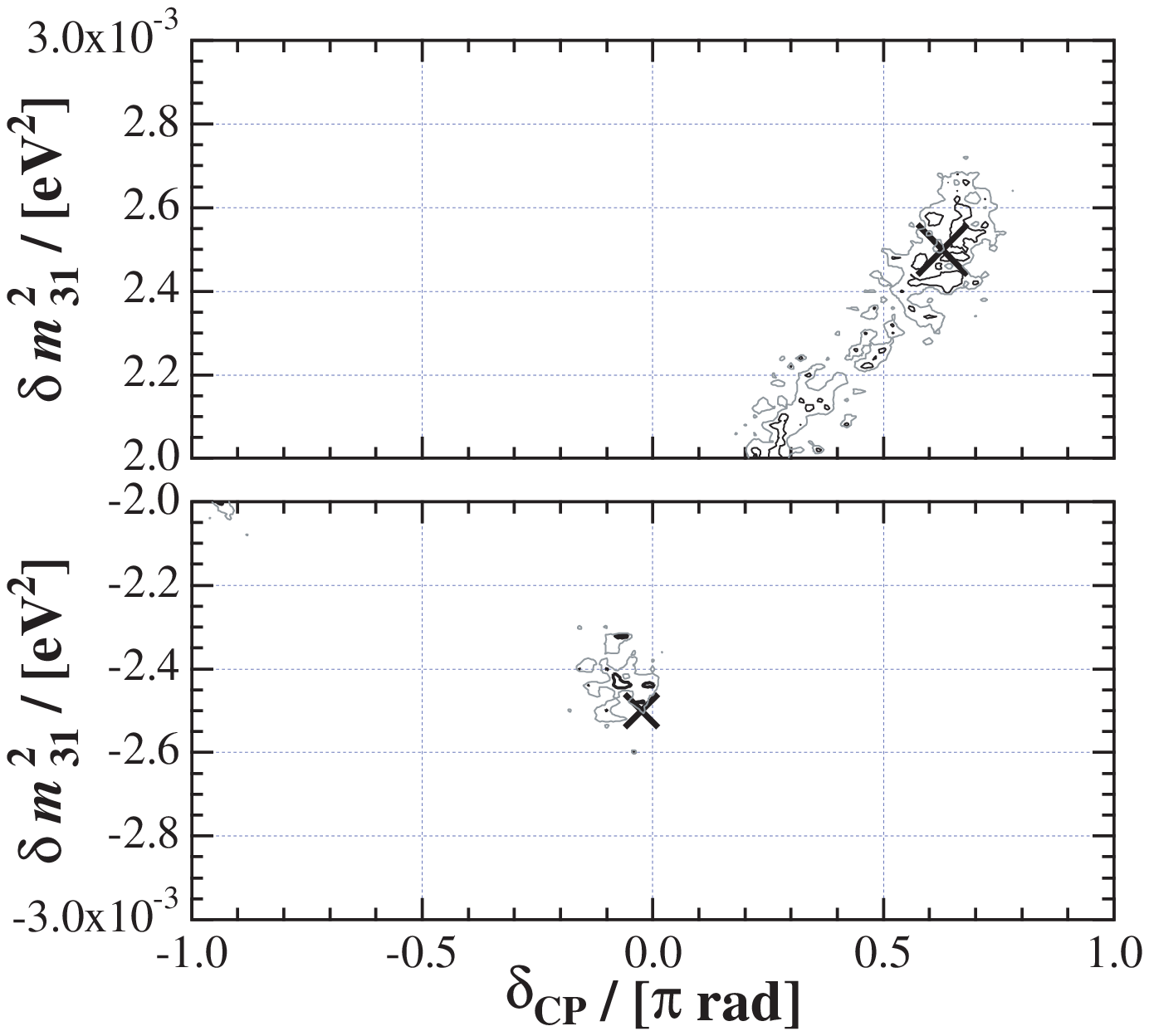}
  }
  \caption{ %
    Allowed regions for 68.3\% (gray curve) and 95\% (black curve)
    confidence levels obtained from an example $\chi^{2}$ analysis of
    the $\nu_{\mu} \to \nu_{\textrm{e}}$ appearance events between
    $0.7 \, \mathrm{GeV}$ and $4.1 \, \mathrm{GeV}$.
    The large ``$\times$'' symbol in the top figure indicates the true
    parameter values of our choice and the small one in the bottom
    indicates the peak-matching partner of the true values as
    explained in the text.
  }
  \label{fig:chi2}
}
\end{figure}
We confirm the validity of the intuitive discussion above by carrying
out a quantitative comparison of the two spectra shown in
Fig.~\ref{fig:appearance-probs}(c), following the analysis given in
Sec.~III~A of Ref.~\citen{Koike:2005dk}.
Our analysis assumes the neutrino beam flux shown in
Fig.~\ref{fig:nu-flux}, which is the spectrum of the wide-band
neutrino beam expected from the upgraded Alternate Gradient
Synchrotron at Brookhaven National Laboratory in the
U.S.A.\cite{Diwan:2003bp} \
This flux is suitable for the present example where we continue to use
$L = 700 \, \mathrm{km}$, since its maximal energy $E \simeq (1 \,
\textrm{--} \, 2) \, \mathrm{GeV}$ coincides with the first-peak
energy of the oscillation probability at this distance for $|\delta
m^{2}_{31}| \simeq (2 \, \textrm{--} \, 3) \times 10^{-3} \,
\mathrm{eV^{2}}$.
We also assume a water \v{C}erenkov detector with a fiducial mass of
$500 \, \mathrm{kt}$ and a data acquisition time of $5 \times 10^{7}
\, \mathrm{s}$.
Other details of the setups are the same as in
Ref.~\citen{Koike:2005dk}.
The analysis is briefly outlined as follows.
We fix the set of ``true'' parameter values as
\begin{math}
  (\delta m^{2}_{31}, \delta)
  = (+2.5 \times 10^{-3} \, \mathrm{eV}^{2}, 0.63 \pi)
\end{math}
taken from Fig.~\ref{fig:appearance-probs}(c), along with the values
in Eq.~(\ref{eq:example-params}).
The ``test'' values of $\delta$ and $\delta m^{2}_{31}$ are varied
over the parameter space while other parameters are fixed to their
true values.
We generate an energy spectrum for this test value and check its
consistency with the true value via a $\chi^{2}$ goodness-of-fit
analysis.
The test values that pass this check constitute the allowed region.
Figure \ref{fig:chi2} shows the allowed regions with 68.3\% and 95\%
confidence levels obtained from the $\chi^{2}$ analysis over the range
$0.7 \, \mathrm{GeV} < E < 4.1 \, \mathrm{GeV}$.
The top graph is for the normal hierarchy, and the bottom for the
inverted.
The large ``$\times$'' symbol on the top graph indicates the true
values of the parameters, and the allowed region extends around it due
to statistical error, systematic error, and the background.
Also in the bottom graph is the allowed region, which indicates the
presence of a hierarchy degeneracy.
The small ``$\times$'' symbol located at
\begin{math}
  (\delta m^{2}_{31}, \delta) = 
  (-2.5 \times 10^{-3} \, \mathrm{eV}^{2}, -0.02\pi)
\end{math}
indicates the peak-matching partner of the true value of our choice
shown in Fig.~\ref{fig:appearance-probs}(c) as
$-0.02\pi \equiv 1.98\pi \;\mathrm{mod}\; 2\pi$.
Note that it falls just within the allowed region with the wrong
hierarchy, indicating the degeneracy of the two sets of parameter
values.

\section{Degeneracy in light of the peak loops}
\label{sec:Peak-loops-in-action}
We showed in the previous section that the degeneracy follows from
the two parameter sets which bring the oscillation peak to the same
position.
We employ the peak loops in this section to present a simple
understanding of the presence of the degeneracy.

\subsection{Emergence of degeneracy: case studies}
\begin{wrapfigure}[20]{l}{\halftext}
  \centerline{
    \includegraphics[width=65mm]{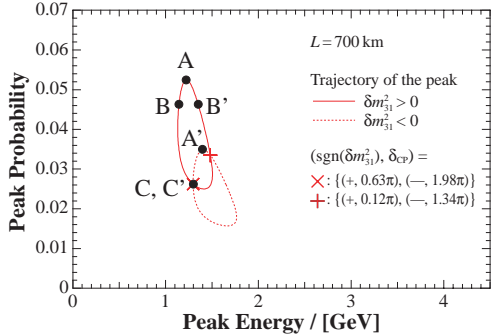}
  }
  \caption{
      Trajectories of the first oscillation peak for the baseline length of
      $700 \, \mathrm{km}$.
      The oscillation parameters in Eq.~(\ref{eq:example-params}) are
      adopted.
      The solid curve is for the normal hierarchy and the dotted one
      for the inverted.
      The ``$\times$'' and ``$+$'' symbols denote the intersections
      of the trajectories.
      The points of $\mathrm{A}$, $\mathrm{A'}$, $\mathrm{B}$,
      $\mathrm{B'}$, $\mathrm{C}$, and $\mathrm{C'}$ are referred to in
      the text.
    }
    \label{fig:loops-700km}
\end{wrapfigure}
We explain how we can read the presence and absence of the
hierarchy degeneracy from the peak loops introduced in the previous
section.
In Fig.~\ref{fig:loops-700km}, we reproduce the pair of peak loops
presented in Fig.~ \ref{fig:appearance-probs}(c).
The upper loop for $\delta m^{2}_{31} > 0$ and the lower loop for
$\delta m^{2}_{31} < 0$ intersect at two points, each of which gives a
pair of values of $(\mathrm{sgn} \, \delta m^{2}_{31}, \delta)$;
in this example the two pairs are
\begin{equation}
\begin{aligned}
  & \bigl\{ (+, 0.63\pi) \, , (-, 1.98 \pi) \bigr\} \, , \quad
  \\
  & \bigl\{ (+, 0.12\pi) \, , (-, 1.34 \pi) \bigr\} \, .
\label{eq:ex-degenerate-pairs}
\end{aligned}
\end{equation}
The true values of $(\mathrm{sgn} \, \delta m^{2}_{31}, \delta)$
provided by Nature give an oscillation spectrum whose first peak
falls on some point on the two loops.
We can correctly determine these values by identifying this point
through experiments.
Sufficient precision and accuracy are necessary to put the
determination into practice.
Even well-controlled experiment, however, can fail to determine
uniquely the parameter values if the true peak falls right upon an
intersection of the loops, giving two possible sets of values as seen
in Eq.~(\ref{eq:ex-degenerate-pairs}).
We demonstrate how degeneracies emerge in three typical cases where
the true peak is at the points A, B, and C in
Fig.~\ref{fig:loops-700km}.

Case (A): %
The true hierarchy is normal and the true oscillation peak is at the
point $\mathrm{A}$ in Fig.~\ref{fig:loops-700km}.
Owing to the vertical separation of the two loops, the sign of $\delta
m^{2}_{31}$ is determined as positive once the appearance probability
is determined precisely enough to distinguish the point A from the
point $\mathrm{A'}$.
The value of $\delta$ is restricted to a single allowed region around
the value for A and the size of this region depends on the
experimental precision and accuracy.

Case (B): %
The true hierarchy is normal and the true oscillation peak is at the
point $\mathrm{B}$ in Fig.~\ref{fig:loops-700km}.
The value of $\delta$ is constrained to a single allowed region as in
the previous case if the energy resolution of the experiment is high
enough to distinguish the points $\mathrm{B}$ and $\mathrm{B'}$ on the
parallel sides of the loop.
If, on the other hand, the resolution is not sufficient, the values of
$\delta$ for $\mathrm{B}$ and for $\mathrm{B'}$ become
indistinguishable and the allowed region will extend around these two
values.

Case (C): %
The true hierarchy is normal, and the true oscillation peak is at the
point $\mathrm{C}$, which is the intersection of the two loops.
There exists another value of $\delta$ that brings the oscillation
peak to the same point $\mathrm{C'}$ but with the inverted hierarchy,
and we are thus led to a hierarchy degeneracy.
Here we regard the points $\mathrm{C}$ and $\mathrm{C'}$ to be on the
upper and lower loops, respectively, in spite of their coincident
positions.
The allowed region consists of two separate parts which extend around
the parameters for $\mathrm{C}$ and for $\mathrm{C'}$.

\subsection{Degeneracy in varying the baseline length}
\label{subsec:degeneracy-and-length}
\begin{wrapfigure}{r}{\halftext}
  \centerline{
    \includegraphics[width=65mm]{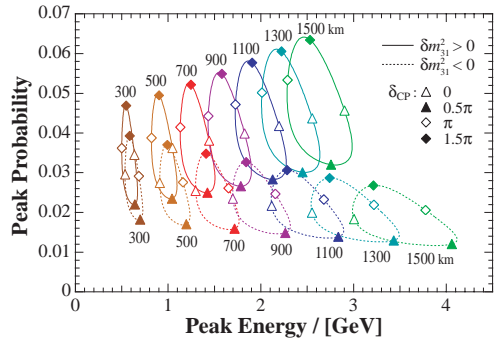}
   }
    \caption{
      Trajectories of the first oscillation peak for
      baseline lengths between $300 \, \mathrm{km}$ and
      $1500 \, \mathrm{km}$.
      The oscillation parameters in Eq.~(\ref{eq:example-params})
      are adopted.
      The solid curve is for the normal hierarchy and the dotted one
      for the inverted.
    }
    \label{fig:nu_trajectory}
\end{wrapfigure}
Next we examine how a pair of loops move and become distorted as we
vary the baseline length of the neutrino oscillation experiments.
We present in Fig.~\ref{fig:nu_trajectory} peak loops obtained as we
vary the baseline length from $300 \, \mathrm{km}$ to $1500 \,
\mathrm{km}$, while fixing the other parameters to the example values
given in Eq.~(\ref{eq:example-params}).
For a relatively short baseline ($L \sim 300 \mathrm{km}$ in our
example),
the loop for the normal hierarchy (the ``normal loop'') and 
         for the inverted hierarchy (the ``inverted loop'')
have similar shapes and overlap each other significantly.
As the baseline becomes longer, the normal loop and the inverted loop
move upward and downward, respectively, and they diverge due to the
matter effect.
The inverted loop is at the same time appreciably stretched in the
$E$-direction and flattened in the $P$-direction.
The two loops are seen to become disjoint at a certain baseline length.
This critical length $L_{\textrm{crit}}$ for the present parameter set
is $L_{\textrm{crit}} \simeq (1100 \,\textrm{--}\, 1300) \,
\mathrm{km}$ as can be read in Fig.~\ref{fig:nu_trajectory}.

From the above observations, we reach following outlook upon the
determination of the hierarchy and the $CP$-violating phase.

The hierarchy is difficult to determine by experiments with a short
baseline length where the pair of loops overlaps considerably.
We expect, however, that the hierarchy can be determined when the true
peak is located at the top part of the upper loop or at the bottom
part of the lower loop.
The determination of the hierarchy becomes easier as the baseline
becomes longer and as the increasing effect of the matter separates
the pair of loops.
When the baseline is longer than the critical length, the condition
given in Eq.~(\ref{eq:peak-matching-cond}) is never satisfied with
opposite hierarchies and the hierarchy can thus be uniquely determined
regardless of the value of $\delta$.

The search for the value of $\delta$ is entangled with that for the
hierarchy.
The hierarchy degeneracy is one obstacle to the determination of
the value of $\delta$ especially when the baseline is short.
Another obstacle is the experimental limitation on the energy
resolution and on the precision and the accuracy of the oscillation
probability.
The loop is narrow in the $E$-direction when the baseline is short,
and the resolution of the energy must be sufficiently high to
distinguish the parallel sides of the loops; otherwise an additional
degeneracy will be introduced.
In contrast, the inverted loop is flattened in the $P$-direction when
the baseline is long.
We thus require precise and accurate measurements of the oscillation
probability to avoid introducing an extra degeneracy.

\section{Analytic study of the peak loops}
\label{sec:Analytic-expressions}
We have illustrated the determination of parameters with the aid of
peak loops and found their position, size, and shape informative.
In this section, we exploit analytic expressions of the oscillation
probability to formulate them in terms of the mass parameters, the
mixing parameters, and the baseline length.
We thereby analyze how a pair of peak loops for $\delta m^{2}_{31}
\gtrless 0$ are distorted and separated as the baseline length
increases.

\subsection{Formulation of the peak loops via the oscillation
  probability formula}
\label{subsec:Peak-loop-formulation}
We derive formulae for the oscillation probability with a perturbative
expansion of the $S$-matrix in terms of $\Delta_{21} \equiv \delta
m^{2}_{21} L / 2E$ and $\Delta_{\textrm{m}} \equiv \sqrt{2}
G_{\textrm{F}} n_{\textrm{e}} L$, where $G_{\textrm{F}} = 1.166 \times
10^{-5} \, \mathrm{GeV}^{-2}$ is the Fermi constant.
The derivation was first worked out to first order to analyze the
effect of $CP$ violation separately from the matter effect.%
\cite{AKS-related} \ 
In the present paper, we calculate the $\nu_{\mu} \to
\nu_{\textrm{e}}$ appearance probability to second order with the
following consideration.
Note that a peak loop collapses to a point when $\delta m^{2}_{21}$
vanishes since $CP$ violation is a three-generation effect.
The size of the peak loop is thus of first order, and its distortion
depending on the baseline length, in which we are interested, is of
second order or higher.
The second-order calculation outlined in Appendix
\ref{app-sec:AKS-approx-deriv} results in lengthy expressions and we
apply an additional simplification that exploits the smallness of
$\theta_{13}$.
We drop $O(\sin^{2} \theta_{13})$ terms in the coefficients of
$\Delta_{21}^{2}$ and $\Delta_{\textrm{m}} \Delta_{21}$, as well as
$O(\sin^{3} \theta_{13})$ terms in that of $\Delta_{\textrm{m}}^{2}$.
Here we take account of the relation $\Delta_{\textrm{m}} >
\Delta_{21}$, which holds for the cases in which we are interested
(see Appendix \ref{app-sec:AKS-approx-validity}).
We then obtain
\begin{subequations}
\begin{equation}
  P(\nu_{\mu} \rightarrow \nu_{\mathrm{e}}, E) 
  = 4l ( A \sin^{2} \Theta + B ) \, ,
\label{eq:Pm2e-aks-order2-body}
\end{equation}
where
\begin{equation}
\begin{split}
  A
  & =
    1 + 2\frac{\Delta_{\textrm{m}}}{\Delta_{31}} (1-2s_{13}^2) 
  - \Delta_{21}\frac{j}{l}\sin\delta
  - \Delta_{21} \frac{\Delta_{\textrm{m}}}{\Delta_{31}} \frac{j}{l}
    \Bigl( \sin\delta + \frac{\Delta_{31}}{2} \cos\delta \Bigr)
  \\ & \quad
  + \frac{\Delta_{21}^2}{2} \frac{j}{l}
    \biggl[ \frac{j}{l}\cos\delta +(1 - 2s_{12}^{2}) \biggr] \cos\delta 
  + 3 \frac{\Delta_{\textrm{m}}^{2}}{\Delta_{31}^{2}} \, ,
\end{split}
\label{eq:Pm2e-aks-order2-A}
\end{equation}
\begin{equation}
\begin{split}
  \Theta
  & =
    \frac{\Delta_{31}}{2} 
  - \frac{\Delta_{\textrm{m}}}{2}(1 - 2s_{13}^{2} )
  + \frac{\Delta_{21}}{2} \Bigl( \frac{j}{l}\cos\delta - s_{12}^{2} \Bigr)
  \\ & \quad
  - \frac{\Delta_{21}}{2} \frac{\Delta_{\textrm{m}}}{\Delta_{31}} \frac{j}{l} 
    \Bigl( \cos\delta + \frac{\Delta_{31}}{2} \sin\delta \Bigr)
    + \frac{\Delta_{21}^2}{2} \frac{j}{l}
    \biggl[ \frac{j}{l}\cos\delta + \frac{1}{2}(1 - 2s_{12}^{2}) \biggr]
    \sin\delta \, ,
\end{split}
\label{eq:Pm2e-aks-order2-Theta}
\end{equation}
and
\begin{equation}
\begin{split}
  B & = \frac{\Delta_{21}^{2}}{4} \frac{j^{2}}{l^{2}} \sin^{2}\delta \, .
\end{split}
\label{eq:Pm2e-aks-order2-B}
\end{equation}
\label{eq:Pm2e-aks-order2}
\end{subequations}
Here
$\Delta_{ij} = \delta m^{2}_{ij} L/2E$,
$l = c_{13}^{2} s_{13}^{2} s_{23}^{2}$, and
$j = c_{13}^{2} s_{13} c_{23} s_{23} c_{12} s_{12}$,
where
$s_{ij} = \sin\theta_{ij}$,
and $c_{ij} = \cos\theta_{ij}$.
The approximation we employed in deriving
Eq.~(\ref{eq:Pm2e-aks-order2}) is suitable especially for a short
baseline, typically for $L < O(10^{3} \, \mathrm{km})$; see Appendix
\ref{app-sec:AKS-approx-validity}.
The oscillation probability for anti-neutrinos $P(\bar{\nu}_{\mu}
\rightarrow \bar{\nu}_{\mathrm{e}}, E) $ is obtained by changing the
signs of $\delta$ and $\Delta_{\textrm{m}}$.

We obtain the peak energy, which gives the local maxima of the
oscillation probability of Eq.~(\ref{eq:Pm2e-aks-order2}), as
\begin{equation}
\begin{split}
  E_{\mathrm{peak}, n}
  & =
  \frac{|\delta m^{2}_{31}| L}{2\Pi}
  \Biggl\{
    \biggl[
          1
      \mp \Delta_{\textrm{m}}
          \frac{1}{\Pi} \Bigl( 1 - \frac{4}{\Pi^{2}} \Bigr)
          \bigl(1 - 2 s_{13}^{2} \bigr)
    \\ &
      \mp R s_{12}^{2}
        + \Delta_{\textrm{m}}^{2} \frac{1}{\Pi^{2}}
          \Bigl( 1 - \frac{12}{\Pi^{2}} + \frac{48}{\Pi^{4}} \Bigr)
        - R^{2} \frac{1}{2} \Bigl( 1 - \frac{4}{\Pi^{2}} \Bigr)
          \frac{j^{2}}{l^{2}}
    \biggr]
    \\ &
    + R
    \biggl[
      \pm 1
        - \Delta_{\textrm{m}} \frac{1}{\Pi} \Bigl( 1 - \frac{8}{\Pi^{2}} \Bigr)
        - 2 R (1 - 2 s_{12}^{2})
    \biggr]
    \frac{j}{l}
    \cos \delta
    \\ &
    + R \frac{2}{\Pi}
    \biggl[
          1
      \mp \Delta_{\textrm{m}} \frac{\Pi}{4}
          \Bigl( 1 + \frac{8}{\Pi^{2}} - \frac{64}{\Pi^{4}} \Bigr)
      \pm R \frac{\Pi^{2}}{4} (1 - 2 s_{12}^{2})
    \biggr]
    \frac{j}{l}
    \sin \delta
    \\ &
    - R^{2}
      \frac{3}{2} \Bigl( 1 + \frac{4}{3} \frac{1}{\Pi^{2}} \Bigr)
      \frac{j^{2}}{l^{2}}
      \cos 2 \delta
    \pm R^{2} \frac{\Pi}{2} \frac{j^{2}}{l^{2}} \sin 2 \delta
  \Biggr\} \, ,
\end{split}
\label{eq:PeakE_2nd}
\end{equation}
where $n = 0, 1, 2, \cdots$ is the peak index, $\Pi \equiv (2n + 1)\pi$, $R
\equiv \delta m^{2}_{21}/|\delta m^{2}_{31}|$, and the top of the double sign
is for $\delta m^{2}_{31} > 0$ and the bottom for $\delta m^{2}_{31} <
0$.
The oscillation probability at the peak energy is given by
\begin{equation}
\begin{split}
  P_{\textrm{peak}, n}
  & =
  4l
  \Biggl\{
    \biggl[
          1 
      \pm \Delta_{\textrm{m}} \frac{2}{\Pi} \bigl( 1 - 2 s_{13}^{2} \bigr)
        + R^{2} \frac{3}{8} \Pi^{2}
          \Bigl( 1 + \frac{4}{3} \frac{1}{\Pi^{2}} \Bigr)
          \frac{j^{2}}{l^{2}}
    \\ &
        + \Delta_{\textrm{m}}^{2} \frac{1}{\Pi^{2}}
          \Bigl( 1 + \frac{4}{\Pi^{2}} \Bigr)
    \biggr]
    -
    R \frac{\Pi^{2}}{2}
    \biggl[
        \Delta_{\textrm{m}} \frac{1}{\Pi} \Bigl( 1 - \frac{4}{\Pi^{2}} \Bigr)
      - R (1 - 2 s_{12}^{2})
    \biggr]
    \frac{j}{l} \cos \delta
    \\ &
    -
    R \Pi
    \biggl[
          1
      \pm \Delta_{\textrm{m}} \frac{2}{\Pi} \Bigl( 1 - \frac{2}{\Pi^{2}} \Bigr)
      \pm R s_{12}^{2}
    \biggr]
    \frac{j}{l} \sin \delta
    \\ &
    +
    R^{2} \frac{\Pi^{2}}{8} \Bigl( 1 - \frac{4}{\Pi^{2}} \Bigr)
    \frac{j^{2}}{l^{2}}
    \cos 2 \delta
    \pm
    R^{2} \frac{\Pi}{2} \frac{j^{2}}{l^{2}} \sin 2 \delta
  \Biggr\} \, .
\end{split}
\label{eq:PeakP_2nd}
\end{equation}
A peak loop is obtained when we keep track of $(E_{\textrm{peak}, n},
P_{\textrm{peak}, n})$ as we vary $\delta$ with other parameters
fixed.
\begin{figure}[t]
\parbox{\halftext}{
  \centerline{
    \includegraphics[width=65mm]{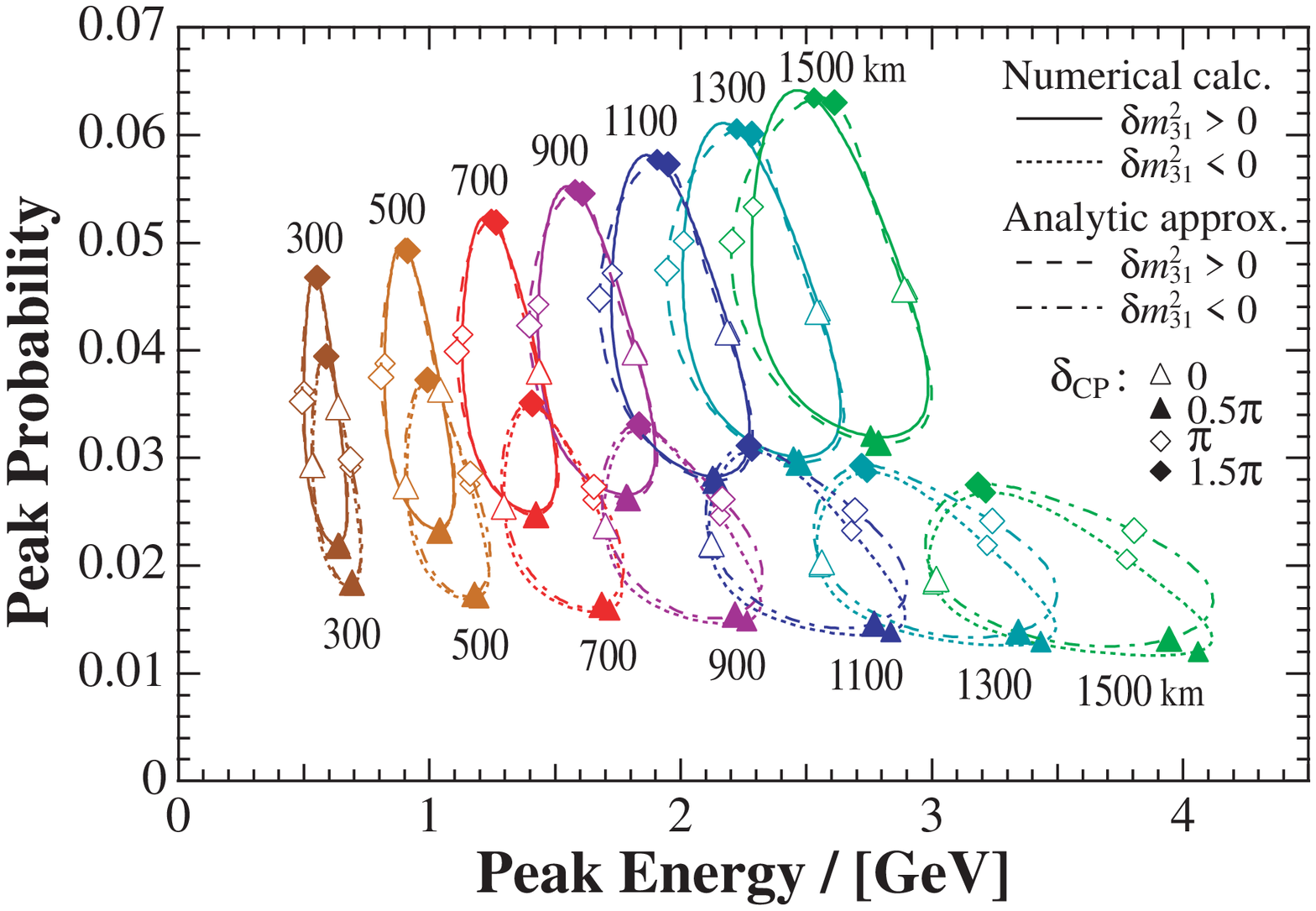}
   } 
    \caption{
      Trajectories of the first oscillation peak for baseline
      lengths between $300 \, \mathrm{km}$ and $1500 \, \mathrm{km}$
      obtained from the numerical calculation (the solid curve for the
      normal hierarchy and the dotted for the inverted) and analytic
      approximations (the dashed curve for the normal hierarchy and the
      dash-dotted for the inverted).
      The oscillation parameters in Eq.~(\ref{eq:example-params}) are
      adopted.
    }
    \label{fig:numerical-vs-analytic}
}
\hfill
\parbox{\halftext}{
  \centerline{
    \includegraphics[width=65mm]{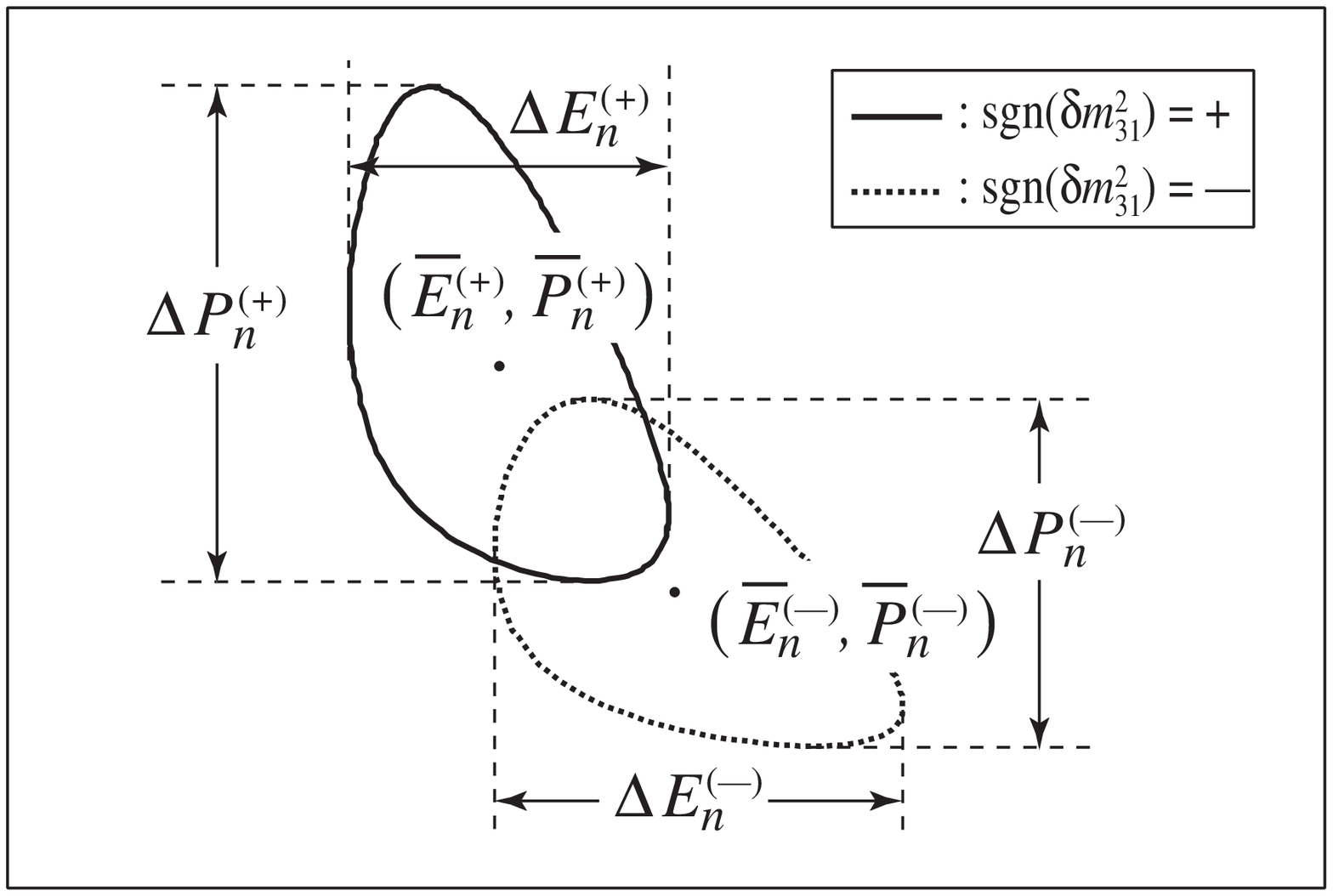}
  }  
    \caption{
      Notations to designate the central position, the width, and
      the height of the trajectories of the $(n + 1)$-th oscillation
      peak.
    }
    \label{fig:loop-notations}
}
\end{figure}
We present in Fig.~\ref{fig:numerical-vs-analytic} the peak loops for
the first peak ($n = 0$) obtained from Eqs.~(\ref{eq:PeakE_2nd}) and
(\ref{eq:PeakP_2nd}) and compare them with the numerical results to
confirm good agreement.

The basic properties of the peak loop are its position, size, and
shape, and we quantify them using its central position, width, and
height, respectively, as follows.
The terms independent of $\delta$ in the braces of
Eqs.~(\ref{eq:PeakE_2nd}) and (\ref{eq:PeakP_2nd}) give the average
values of $E_{\mathrm{peak}, n}$ and $P_{\mathrm{peak}, n}$ over a
cycle of $\delta$ and can be regarded as the center of a peak loop.
The central position up to first order thus reads
\begin{equation}
\begin{split}
  \bigl( \overline{E}_{n}^{(\pm)}, \overline{P}_{n}^{(\pm)} \bigr)
  =
  \Biggl(
  &
    \frac{|\delta m^{2}_{31}| L}{2\Pi}
    \biggl[
          1
      \mp \Delta_{\textrm{m}} \frac{1}{\Pi}
          \Bigl( 1 - \frac{4}{\Pi^{2}} \Bigr)
          \bigl( 1 - 2s_{13}^{2} \bigr)
      \mp  R s_{12}^{2}
    \biggr]
    \, , \;
    \\ &
    4l
    \biggl[
      1 \pm \Delta_{\textrm{m}} \frac{2}{\Pi}
            \bigl( 1 - 2s_{13}^{2} \bigr)
    \biggr]
  \Biggr) \, ,
\end{split}
\label{eq:loop_center_nu}
\end{equation}
where the double sign is the same as in Eqs.~(\ref{eq:PeakE_2nd}) and
(\ref{eq:PeakP_2nd}); see Fig.~\ref{fig:loop-notations}.
The terms that depend on $\delta$ in Eqs.~(\ref{eq:PeakE_2nd}) and
(\ref{eq:PeakP_2nd}) account for the size and the shape of a peak
loop.
The width of a loop $\Delta E_{n}^{(\pm)}$ and its height $\Delta
P_{n}^{(\pm)}$ can be estimated by taking the difference of the
maximum and minimum values of $E_{\textrm{peak}, n}$ and
$P_{\textrm{peak}, n}$ as functions of $\delta$ (see
Fig.~\ref{fig:loop-notations}).
They are calculated up to second order as
\begin{subequations}
\begin{align}
  &
  \begin{aligned}
  \Delta E_{n}^{(\pm)}
  =
  &
  |\delta m^{2}_{31}| L R \frac{1}{\Pi} \sqrt{ 1 + \frac{4}{\Pi^{2}} }
  \frac{j}{l}
  \\ & \times
  \biggl[ 1 \mp \Delta_{\textrm{m}} \frac{2}{\Pi}
                \frac{ 1 - 32/\Pi^{4} }{ 1 + 4/\Pi^{2} }
            \mp R \frac{1}{1 + 4/\Pi^{2}} \bigl( 1 - 2s_{12}^{2} \bigr)
  \biggr] \, ,
  \end{aligned}
  \label{eq:Delta_E_nu}
  \\
  &
  \Delta P_{n}^{(\pm)}
  =
  8R\Pi j
  \biggl[
    1 \pm \Delta_{\textrm{m}} \frac{2}{\Pi} \Bigl( 1 - \frac{2}{\Pi^{2}} \Bigr)
      \pm Rs_{12}^{2}
  \biggr] \, .
  \label{eq:Delta_P_nu}
\end{align}
\label{eq:Delta_EP_nu}
\end{subequations}

We remark on the magnitude of the correction terms appearing in
Eqs.~(\ref{eq:loop_center_nu}) and (\ref{eq:Delta_EP_nu}).
The correction terms include the matter effect proportional to
$\Delta_{\textrm{m}}$ and the three-generation effect proportional to
$R$.
Of these two, the three-generation effect is even smaller than the
matter effect for a typical long baseline experiments.
Evaluation using the example values given in
Eq.~(\ref{eq:example-params}) is sufficient to verify this with an
order-of-magnitude estimation, which yields
\begin{equation}
  \Delta_{\textrm{m}} \frac{1}{\Pi}
  \sim 0.2 \biggl( \frac{L}{1000 \, [\mathrm{km}]} \biggr)
  \quad \textrm{for $n = 0$} \, ,
  \hspace{5mm}
  R(1 - 2s_{12}^{2}) \sim R s_{12}^{2} \sim 0.01 \, .
\label{eq:correction-estimations}
\end{equation}

The expressions for anti-neutrinos are obtained %
from Eqs.~(\ref{eq:loop_center_nu}) and (\ref{eq:Delta_EP_nu}) by
simply flipping the signs of $\Delta_{\textrm{m}}$:
\begin{equation}
\begin{split}
  \bigl( \overline{E}_{n}^{(\pm)}, \overline{P}_{n}^{(\pm)} \bigr)
  =
  \Biggl(
  &
    \frac{|\delta m^{2}_{31}| L}{2\Pi}
    \biggl[
          1
      \pm \Delta_{\textrm{m}} \frac{1}{\Pi} 
          \Bigl( 1 - \frac{4}{\Pi^{2}} \Bigr) (1 - 2s_{13}^{2})
      \mp  R s_{12}^{2}
    \biggr]
    \, , \,
  \\ &
    4l
    \biggl[
          1
      \mp \Delta_{\textrm{m}} \frac{2}{\Pi} (1 - 2s_{13}^{2})
    \biggr]
  \Biggr)
  \, ,
\end{split}
\label{eq:loop_center_antinu}  
\end{equation}
\begin{subequations}
\begin{align}
  &
  \begin{aligned}
    \Delta E_{n}^{(\pm)}
    =
    &
    |\delta m^{2}_{31}| L R \frac{1}{\Pi}
    \sqrt{ 1 + \frac{4}{\Pi^{2}} } \frac{j}{l}
    \\ &
    \times \biggl[
      1 \pm \Delta_{\textrm{m}} \frac{2}{\Pi}
      \frac{ 1 - 32/\Pi^{4} }{ 1 + 4/\Pi^{2} }
      \mp R \frac{1}{1 + 4/\Pi^{2}} \bigl( 1 - 2s_{12}^{2} \bigr)
    \biggr] \, ,
  \end{aligned}
  \label{eq:Delta_E_nubar}
  \\
  &
  \Delta P_{n}^{(\pm)}
  =
  8R\Pi j
  \biggl[
    1 \mp \Delta_{\textrm{m}} \frac{2}{\Pi} \Bigl( 1 - \frac{2}{\Pi^{2}} \Bigr)
      \pm Rs_{12}^{2}
  \biggr] \, .
\label{eq:Delta_P_nubar}
\end{align}
\label{eq:Delta_EP_nubar}
\end{subequations}
Equations (\ref{eq:loop_center_antinu}) and (\ref{eq:Delta_EP_nubar})
have an interesting relation to (\ref{eq:loop_center_nu}) and
(\ref{eq:Delta_EP_nu}).
The former pair for anti-neutrinos is equivalent to the latter for
neutrinos with flipped hierarchy up to the correction terms
proportional to $R s_{12}^{2}$ or $R(1 - 2s_{12}^{2})$, which are
small as shown in Eq.~(\ref{eq:correction-estimations}).

\subsection{Analysis of the peak loops}
\label{subsec:Peak-loop-analysis}

We elevate the observation of peak loops for neutrinos shown by
Fig.~\ref{fig:nu_trajectory} to a systematic analysis by applying the
formulae we developed in the previous subsection.
We also extend our consideration to the anti-neutrino case.

Equation (\ref{eq:loop_center_nu}) reveals the order behind the motion
of the loops for neutrinos in the $E$-$P$ plane as the baseline length
becomes longer.
The leading dependence of $\overline{E}_{n}^{(\pm)}$ on the baseline
length comes from the prefactor $|\delta m^{2}_{31}| L/2E$ and drives
the loops rightward.
Its subleading dependence due to the matter-effect term, which is
proportional to the baseline length through $\Delta_{\textrm{m}}$ and
to $\mathrm{sgn} \, \delta m^{2}_{31}$, provides a correction which
pulls the normal loop back leftward and gives the inverted loop another
push rightward.
The other subleading term $Rs_{12}^{2}$ of $\overline{E}_{n}^{(\pm)}$
gives an extra correction whose sign depends on the hierarchy and
whose magnitude depends on the mass parameters and the mixing
parameters but not on the baseline length.
The dependence of $\overline{P}_{n}^{(\pm)}$ on the baseline length
comes from the matter-effect term, which raises the normal loop and
lowers the inverted loop.
As a whole, a pair of normal and inverted loops are driven rightward
while they split vertically, and their alignment tilts
counterclockwise.
These features of the motion of the loops are clearly observed in
Fig.~\ref{fig:nu_trajectory}.

\begin{wrapfigure}{l}{\halftext}
  \centerline{
    \includegraphics[width=65mm]{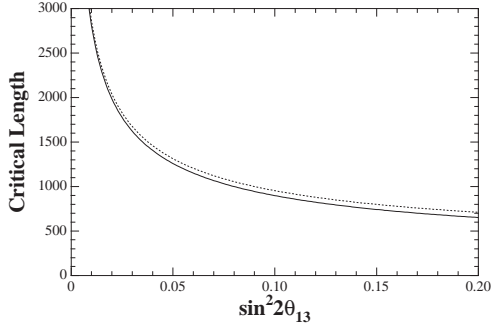}
  }
    \caption{
      The critical length as a function of $\sin^{2} 2\theta_{13}$
      in the first-order approximation, Eq.~(\ref{eq:Lcrit-def}).
      The solid curve is for neutrinos and the dotted curve for
      anti-neutrinos.
      We adopt the values of the oscillation parameters given in
      Eq.~(\ref{eq:example-params}) except for that of
      $\sin^{2} 2\theta_{13}$ in
      Eq.~(\ref{eq:example-params-sin2-2theta13}).
    }
    \label{fig:critical-length}
\end{wrapfigure}
We next analyze the distortion of the loop for neutrinos in terms of
its width and height given in Eq.~(\ref{eq:Delta_EP_nu}).
The expression for $\Delta E_{n}^{(\pm)}$ has an overall factor which
is proportional to the baseline length and accounts for the widening
of the loops.
One of two subleading corrections of $\Delta E_{n}^{(\pm)}$ is the
matter-effect term, which adds a dependence on the baseline length.
The other term also gives a correction, but does not give any
additional dependence on the baseline length.
Both of these terms depend on the hierarchy, and decelerate the widening
of the normal loop and accelerate that of the inverted loop.
The dependence of $\Delta P_{n}^{(\pm)}$ on the baseline length is due
to the subleading matter-effect term, which vertically stretches the
normal loop and compresses the inverted loop, and the other correction
term supplements its effect.
All these features of the loops are evident in
Fig.~\ref{fig:nu_trajectory}.

Now that the motion and the distortion of the loops for the neutrinos
have been analyzed, those for anti-neutrinos are easy to derive owing
to the discussion below Eq.~(\ref{eq:Delta_EP_nubar}) in the previous
subsection.
A loop for anti-neutrinos moves with distortion just as that for
neutrinos with the opposite hierarchy, except for the small
contribution of the three-generation effect.

The critical baseline length mentioned in the previous section also
can be analyzed with our formulation.
The critical length is defined as the maximum length giving the
intersection of the pairing loops with opposite hierarchies.
It is calculated in the first-order approximation as
\begin{equation}
\begin{split}
  L_{\textrm{crit}}
  = &
  \frac{1}{\sqrt{2} G_{\textrm{F}} n_{\textrm{e}}}
  \frac{\Pi}{ 1 - \dfrac{12}{\Pi^{2}} + \dfrac{64}{\Pi^{4}} }
  R
  \frac{c_{23} c_{12} s_{12}}{s_{13} s_{23}}
  \frac{1}{1 - 2 s_{13}^{2}}
  \\ \times &
  \Biggl[
    \sqrt{
      1 - \dfrac{12}{\Pi^{2}} + \dfrac{64}{\Pi^{4}}
        - \frac{4}{\Pi^{2}}
          \Bigl( \frac{s_{13} s_{23} s_{12}}{c_{23} c_{12}} \Bigr)^{2}
    }
    \mp  
    \Bigl( 1 - \dfrac{8}{\Pi^{2}} \Bigr)
    \frac{s_{13} s_{23} s_{12}}{c_{23} c_{12}}
  \Biggr] \, ,
\end{split}
\label{eq:Lcrit-def}
\end{equation}
where the double sign is $-$ for neutrinos and $+$ for anti-neutrinos.
The prefactor appearing on the right-hand side of this expression
gives
\begin{math}
    1/\sqrt{2} G_{\textrm{F}} n_{\textrm{e}}
  = 5.17 \times 10^{3} \, [\mathrm{km}] \times 
    (\rho / [\mathrm{g \, cm^{-3}}])^{-1}
  = 2.0 \times 10^{3} \, [\mathrm{km}] \cdot
    (\rho / [2.6 \, \mathrm{g \, cm^{-3}}])^{-1}.
\end{math}
The critical length is inversely proportional to $\sin \theta_{13}$
owing to the factor $c_{23} c_{12} s_{12}/s_{13} s_{23}$, with the small
correction of $O(\theta_{13})$.
It is particularly important that this dependence can make the
critical length very long, since the lower bound on the value of $\sin
\theta_{13}$ is unknown at present.
Figure \ref{fig:critical-length} shows the dependence of the critical
length on $\sin^{2} 2\theta_{13}$ for the first peak, where the other
parameters are fixed to the values given in
Eq.~(\ref{eq:example-params}).
The critical length for anti-neutrinos is slightly longer than that
for neutrinos due to the small correction proportional to
$s_{13}s_{23}s_{12}/c_{23}c_{12}$.
From this graph, we find $L_{\textrm{crit}} = 1150 \, \mathrm{km}$ for
neutrinos and $L_{\textrm{crit}} = 1200 \, \mathrm{km}$ for
anti-neutrinos at $\sin^{2} 2\theta_{13} = 0.06$.
This graph is qualitatively consistent with Fig.~5 in
Ref.~\citen{Barger:2001yr}, which is obtained from their bi-channel
analysis with the energy fixed around a peak at $E = |\delta
m^{2}_{31}| L / 2 \pi$.
It is remarkable that the following straightforward analysis of the
loops reproduces the main part of Eq.~(\ref{eq:Lcrit-def}), which is
tedious to derive, and clarifies the origin of the dependence of the
critical length upon the mass parameters and the mixing parameters.
We compare the separation between the loops with their size to obtain
a condition for their disentanglement.
The separation of the two loops is estimated using the difference
between the respective values of $\overline{E}^{(\pm)}$ and
$\overline{P}^{(\pm)}$, while the size of each loop is measured by its
width and height.
We consider the ratio of the separation to the size to compare the
two, and define a normalized separation vector as
\begin{equation}
\begin{split}
   &
   \Biggl( 
      \frac{\overline{E}_{n}^{(+)} - \overline{E}_{n}^{(-)}}
           {\bigl[ \Delta E_{n}^{(+)} + \Delta E_{n}^{(-)} \bigr]/2}, \ 
      \frac{\overline{P}_{n}^{(+)} - \overline{P}_{n}^{(-)}}
           {\bigl[ \Delta P_{n}^{(+)} + \Delta P_{n}^{(-)} \bigr]/2}
    \Biggl)
    \\ &  \hspace{30mm}
    \simeq
    \frac{\Delta_{\textrm{m}}}{R} \frac{l}{j} (1-2s_{13}^{2}) 
    \biggl(
       -\frac{1}{\Pi} \frac{1-4/\Pi^{2}}{\sqrt{1+4/\Pi^{2}}}, \ 
        \frac{2}{\Pi^{2}}
    \biggr) \, ,
\end{split}
\label{eq:Ebar-Pbar}
\end{equation}
where we have omitted small terms such as $Rs_{12}^{2}$ and $R(1 -
s_{12}^{2})$ on the right-hand side for simplicity.
The magnitude of this vector is given by
\begin{equation}
\begin{split}
   S(L)
   & \simeq
   \frac{\Delta_{\textrm{m}}}{R} \frac{l}{j} (1 - 2s_{13}^{2})
   \frac{1}{\Pi}
   =
   \frac{\sqrt{2} G_{\textrm{F}} n_{\textrm{e}}}{(2n + 1)\pi} L
   \frac{1}{R} \frac{l}{j} (1 - 2s_{13}^{2})
   \, ,
 \end{split}
 \label{eq:separation}   
\end{equation}
which is proportional to $L$ and parametrizes the configuration of the
paired loops, \textit{i.e.} intersected, tangential, or disjoint.
The normalization of Eq.~(\ref{eq:Ebar-Pbar}) is chosen so that the
separation at the critical length $S_{\textrm{crit}} \equiv
S(L_{\textrm{crit}})$ becomes unity when the loops are two similar
ellipses whose major axes are parallel.
We write the critical length from Eq. (\ref{eq:separation}) as
\begin{equation}
\begin{split}
   L_{\textrm{crit}}
   & =
   \frac{(2n + 1)\pi}{\sqrt{2} G_{\textrm{F}} n_{\textrm{e}}}
   R
   \frac{c_{23}c_{12}s_{12}}{s_{23}} \frac{1}{s_{13}}
   \frac{1}{1 - 2s_{13}^{2}}
   S_{\textrm{crit}} \, .
\end{split}
\label{eq:Lcrit-alternative}
\end{equation}
Equation (\ref{eq:Lcrit-alternative}) approximately reproduces the
dependence on the parameters appearing on the right-hand side of
Eq.~(\ref{eq:Lcrit-def}), assuming that $S_{\textrm{crit}}$ does not
give any extra dependence on the mass parameters and the mixing
parameters.

\section{Toward a resolution of the hierarchy degeneracy}
\label{sec:Resolving-hierarchy-degeneracy}
In this section, we consider how to avoid the hierarchy degeneracy in
determining the value of $\delta$ with long baseline experiments.

\begin{wrapfigure}{l}{\halftext}
  \centerline{
    \includegraphics[width=65mm]{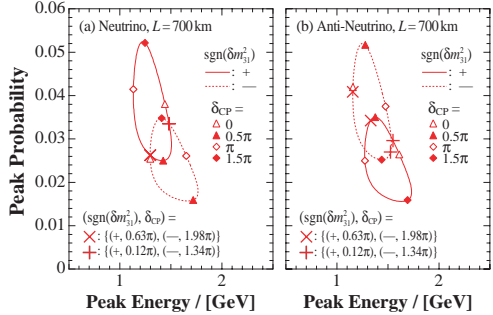}
  }  
    \caption{
      Trajectories of the first peak of (a) the $\nu_{\mu} \to
      \nu_{\textrm{e}}$ probability and (b) the $\bar{\nu}_{\mu} \to
      \bar{\nu}_{\textrm{e}}$ probability for a baseline length of
      $700 \, \mathrm{km}$.
      The solid and dotted curves are for the normal and inverted
      hierarchies, respectively.
      The ``$\times$'' and ``$+$'' symbols in (a) are on the
      intersections of the  peak loops, and the corresponding values
      of the parameters are also shown.
      The symbols in (b) mark the points corresponding to these
      parameter values.
    }
    \label{fig:peak_loops_nu_nubar}
\end{wrapfigure}
A promising approach for determining the value of $\delta$ and the
hierarchy is (1) to carry out an experiment with baseline length
longer than the critical length so that the hierarchy is identified
regardless of the value of $\delta$.
This approach can be made feasible by pinning down the position of
only one peak.
It is nonetheless not free of hurdles: the required baseline length is
typically about $1000 \, \mathrm{km}$ and may be even longer,
depending on the value of $\theta_{13}$.
Performing experiments with such a long baseline is challenging due
to, for instance, the small flux of the neutrino beam and a possible
large ambiguity of matter effects.

Experiments with shorter baselines are more feasible, but leave the
possibility of degeneracy.
To overcome this problem, we can simultaneously make use of two or
more peaks.
We examine the effectiveness of the following three approaches of this
kind:
(2) observing $\bar{\nu}_{\mu} \to \bar{\nu}_{\mathrm{e}}$ appearance
events in addition to $\nu_{\mu} \to \nu_{\mathrm{e}}$ events;
(3) carrying out two or more experiments with different baseline lengths;
and 
(4) carrying out an experiment which has two or more oscillation peaks
within its range of visible neutrino energy.

The second approach employs both neutrinos and anti-neutrinos.
We analyze the merit of this approach with
Fig.~\ref{fig:peak_loops_nu_nubar}, which presents two pairs of peak
loops, one (a) for neutrinos and the other (b) for anti-neutrinos,
with the values of Eq.~(\ref{eq:example-params}) and $L = 700 \,
\mathrm{km}$.
The loops in Figs.~\ref{fig:peak_loops_nu_nubar}(a) and (b) resemble
each other with opposite assignments of the hierarchy, confirming our
discussion in \S\ref{subsec:Peak-loop-analysis} based on analytic
formulae.
Marked by the ``$\times$'' and ``$+$'' symbols in
Fig.~\ref{fig:peak_loops_nu_nubar}(a) are the two intersections of
the normal and inverted loops.
Each intersection is associated with two sets of $(\mathrm{sgn} \,
\delta m^{2}_{31}, \delta)$ as we gave examples in
Eq.~(\ref{eq:ex-degenerate-pairs}) and corresponds to the presence of
the hierarchy degeneracy.
Turning to Fig.~\ref{fig:peak_loops_nu_nubar}(b), the four parameter
sets considered there correspond to four distinct points on the peak
loops for anti-neutrinos.
Hence the combined analysis of $\nu_{\mu} \to \nu_{\mathrm{e}}$ events
and $\bar{\nu}_{\mu} \to \bar{\nu}_{\mathrm{e}}$ events will be able
to resolve the hierarchy degeneracy, provided that the two ``$\times$''
symbols or the two ``$+$'' symbols are experimentally distinguishable.
In our present example, the two ``$+$'' symbols for anti-neutrinos are
close to each other in the $E$-$P$ plane and are thus difficult to
distinguish.
If indistinguishable, the hierarchy degeneracy will persist even with
the aid of anti-neutrino events.

\begin{figure}[t]
\parbox{\halftext}{
  \centerline{
    \includegraphics[width=65mm]{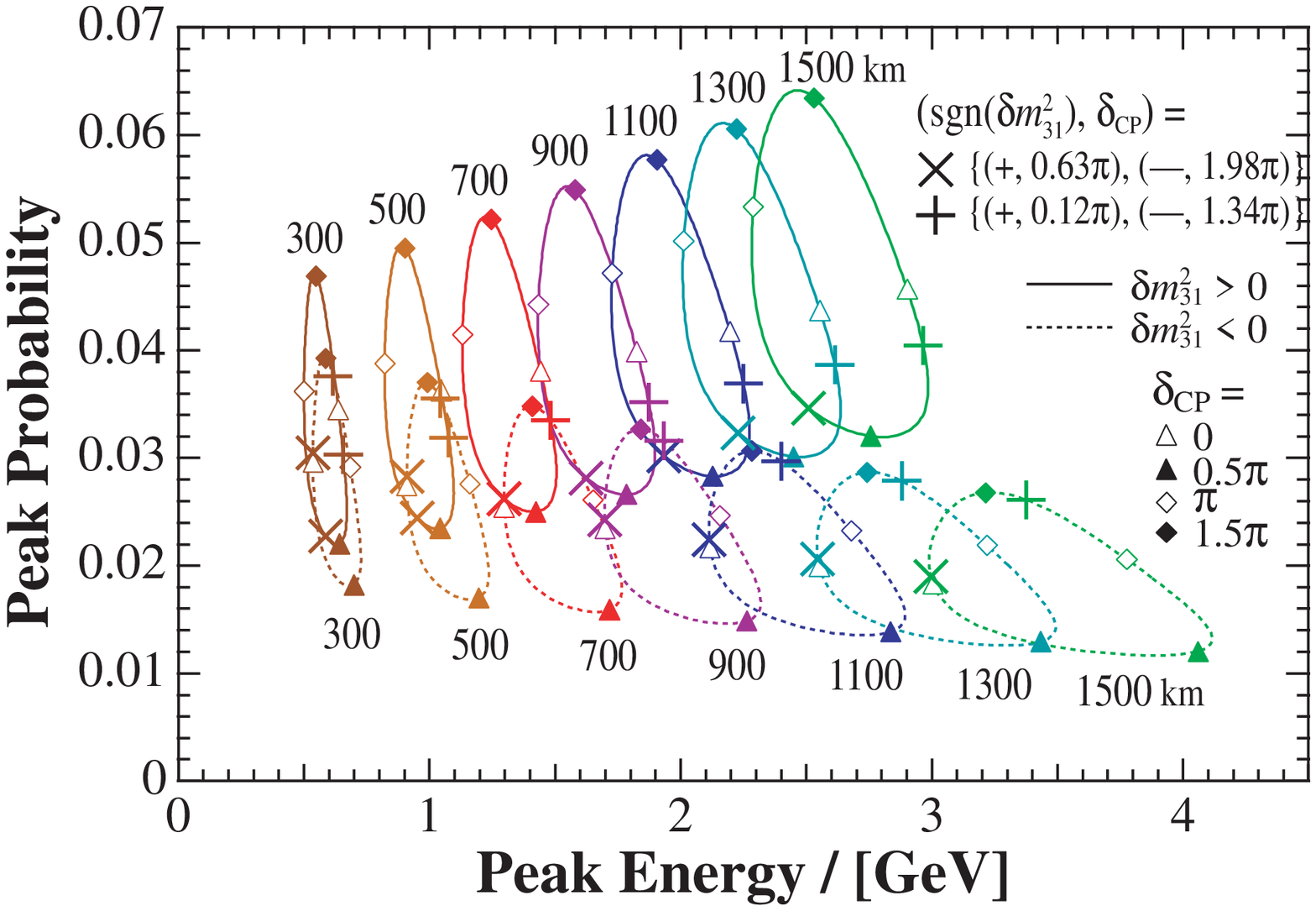}
  }  
    \caption{
      The positions of the first peak for baseline lengths
      between $300 \, \mathrm{km}$ and $1500 \, \mathrm{km}$.
      The ``$\times$'' and ``$+$'' symbols indicate the positions
      for the four sets of parameter values which correspond to
      the two intersections of the peak loops at
      $L = 700 \, \mathrm{km}$,
      and they are plotted on top of the series of peak loops
      reproduced from Fig.~\ref{fig:nu_trajectory}.
      The parameter values given in Eq.~(\ref{eq:example-params}) are
      adopted.
      The solid (dotted) curve is for the normal (inverted)
      hierarchy.
    }
  \label{fig:loops-various-L}
}
\hfill
\parbox{\halftext}{
  \centerline{
    \includegraphics[width=65mm]{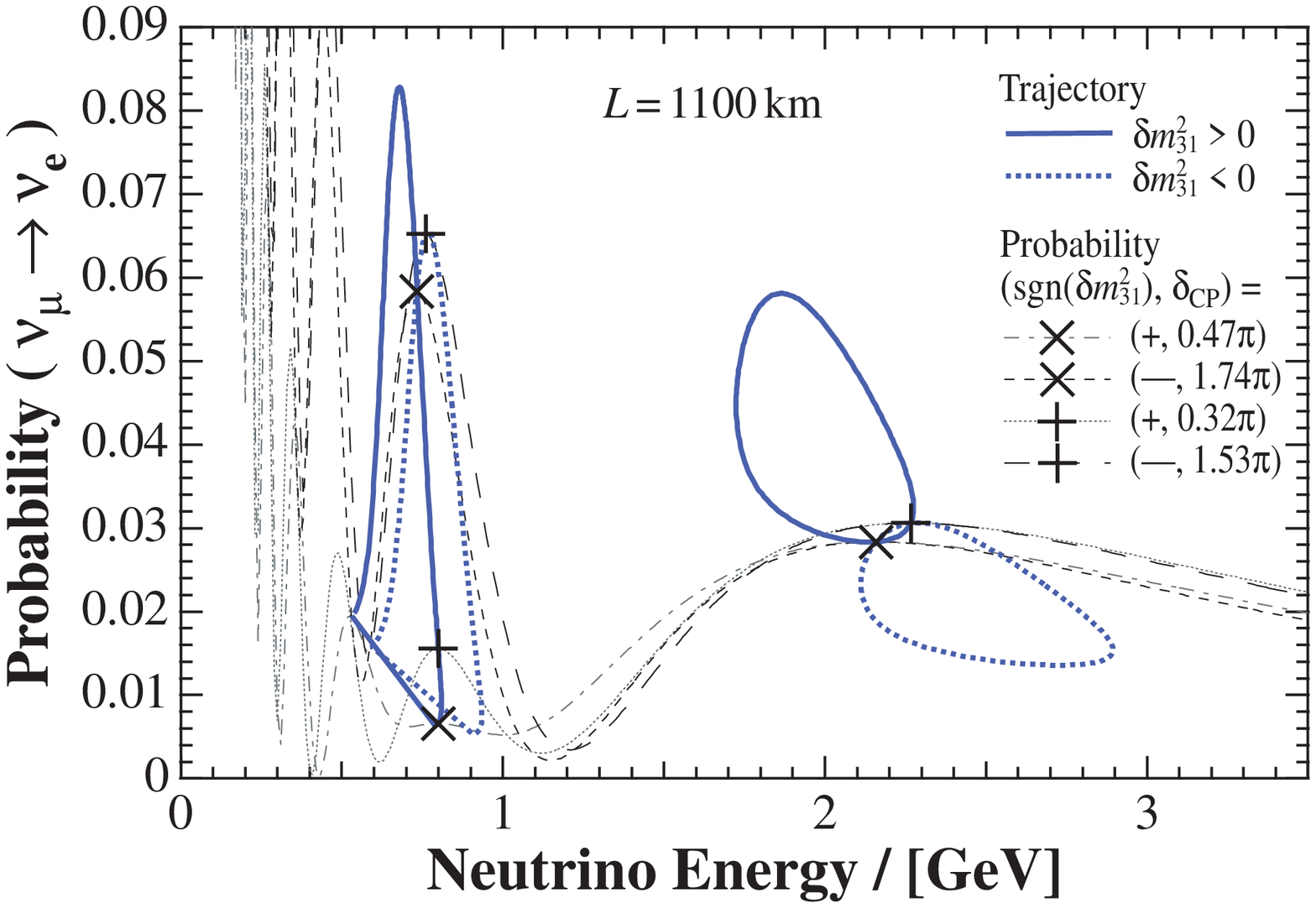}
  }  
    \caption{
      Trajectories of the first and second oscillation peaks for
      $L = 1100 \, \mathrm{km}$.
      The parameter values given in Eq.~(\ref{eq:example-params})
      are adopted.
      The solid (dotted) loop is for the normal (inverted) hierarchy.
      The ``$\times$'' and ``$+$'' symbols correspond to the sets of
      parameter values for which the first peak is at the intersection
      of its trajectories.
      The points for these values are also on the second
      peak loops and are marked by the corresponding symbols.
      The oscillation spectra for the parameter values are overlaid.
    }
    \label{fig:2ndpeak_nu_trajectory}
}
\end{figure}
The third approach makes use of two or more different baseline
lengths.\cite{bib:two-baselines-combo} \ 
We reproduce in Fig.~\ref{fig:loops-various-L} the series of peak loops
shown in Fig.~\ref{fig:nu_trajectory}.
On top of them, we mark the points corresponding to the parameter
values for the intersections of the loops in the case $L = 700 \,
\mathrm{km}$ as we did in Fig.~\ref{fig:peak_loops_nu_nubar}.
Each pair of loops for a fixed baseline length has two ``$\times$''
symbols and two ``$+$'' symbols, which reduce, by definition, to a
single ``$\times$'' symbol and a single ``$+$'' symbol at $L = 700 \,
\mathrm{km}$.
The separation between the two ``$\times$'' symbols as well as between
the two ``$+$'' symbols grows as the baseline length deviates from
$700 \, \mathrm{km}$.
The degeneracy of the two ``$\times$'' symbols or the two ``$+$''
symbols can thus be resolved with an additional experiment with a
different baseline length, where the large difference between the two
baseline lengths is preferable.

The fourth approach exploits other peaks of the appearance probability
in addition to the first one for $n = 0$.\cite{Diwan:2003bp} \
This approach requires a wide-band neutrino beam in order to
accumulate a sufficient number of events at the multiple peaks.
An analysis covering peaks for different values of $n$ is effective to
resolve the degeneracy since each peak has its own properties due to
the factor $\Pi \equiv (2n + 1)\pi$ in Eqs.~(\ref{eq:loop_center_nu})
and (\ref{eq:Delta_EP_nu}).
The peak for $n > 0$, however, is not easy to observe due to its small
energy which is suppressed by the factor of $1/(2n + 1)$.
On the other hand, elongation of the baseline makes the peak energies
higher, bringing both the first and second peaks within the range of
visible energy.
We show in Fig.~\ref{fig:2ndpeak_nu_trajectory} the two sets of loops
traced by the first and second peaks for $L = 1100 \, \mathrm{km}$.
The second-peak loops are stretched in the $P$-direction and have a
significant overlap, showing their noticeable difference from
the first-peak loops.
The ``$\times$'' and ``$+$'' symbols correspond to the values at the
two intersections of the loops for the first peak.
On the second peak loops, the paired ``$\times$'' symbols as well as
the paired ``$+$'' symbol are separated from each other and are
experimentally distinguishable.
The hierarchy degeneracy in the first peak will be thereby removed by
observing the second peak.

\section{Conclusion and discussion}
\label{sec:Conclusion-and-discussions}
We have studied the search for the leptonic $CP$-violating phase
$\delta$ and the neutrino mass hierarchy $\mathrm{sgn} \, \delta
m^{2}_{31}$ through the observation of the $\nu_{\mu} \to
\nu_{\textrm{e}}$ oscillation with long baseline experiments.
The energy spectrum of the $\nu_{\mu} \to \nu_{\textrm{e}}$ appearance
probability is engraved with the values of these parameters, although
not uniquely.
The search for these parameters through this spectrum can lead to
degeneracy when the two spectra corresponding to the two different
parameter values are indistinguishable.
We implemented the presence of degeneracy using the following
criterion: the oscillation probabilities for the two parameter values
are peaked at the same energy and have the same peak probability.
In light of this peak-matching condition, we systematically examined
the presence and absence of degeneracy, especially the hierarchy
degeneracy with regard to the value of $\delta$ and the sign of
$\delta m^{2}_{31}$.
We have also demonstrated the prospects for resolving the degeneracy
using a single experiment and combinations of experiments.

We introduced a looped trajectory traced by the position of the
oscillation peak in the $E$-$P$ plane while varying the value of
$\delta$ over the interval $[0, 2\pi]$.
We have shown that the loop plays a key r\^{o}le in gaining an
intuitive understanding of the parameter degeneracy.
We drew pairs of loops for both hierarchies and investigated their
properties and behavior to obtain an overview of the emergence of the
hierarchy degeneracy and methods to avoid it.
We observed that paired loops with different hierarchies are
completely separated when the baseline is longer than a critical
length, which is proportional to $1/\sin\theta_{13}$ and is typically
about $1000 \, \mathrm{km}$ or longer.
The hierarchy degeneracy can be avoided in experiments with a
baseline length longer than the critical length.

Employing the peak loops, we obtained a general understanding of
experimental capabilities to determine the value of $\delta$ and the
mass hierarchy.
We described four ideas for resolving degeneracy with long baseline
experiments:
using a baseline longer than the critical length, %
employing anti-neutrinos as well as neutrinos, %
using two or more different baseline lengths, %
and carrying out experiments covering two or more oscillation peaks.
We discussed the merits and demerits of each idea in terms of the peak
loops.
We pointed out that the first idea of using a sufficiently long baseline
is simple but is not free of hurdles.
We also found for the second case that employing anti-neutrinos
does not always resolve the hierarchy degeneracy.

The peak loops we introduced to analyze the parameter degeneracy
is as versatile as the trajectories in the bi-channel plots
presented in Ref.~\citen{bib:biprob-plot}.
The two have differences at the same time.
One difference is that our plot employs only a single channel and thus
can be used to explore the capabilities of single-channel
experiments.
Another difference is that we make essential use of the oscillation
peak as a representative of the oscillation spectrum over a finite
energy range, while the bi-channel plot is drawn for an arbitrary
fixed energy or for integrated values over an energy range at one's
convenience.
The significance of the peak position consists in its implication to
the values of the parameters for which we search.
Supported by the peak-matching condition, our method facilitates a
direct prediction of the presence of degeneracy.

We kept the values of the mass parameters and the mixing parameters
fixed except for $\delta$ and $\mathrm{sgn} \, \delta m^{2}_{31}$,
assuming that their values will be determined before the experiments
in question are carried out.
It is possible, however, that they will not be determined with
sufficient precision by that time and that their ambiguities will add
another complication to the searches.%
\cite{bib:ambiguity,Koike:2005dk} \
The peak loops presented in this paper also offer an intuitive
understanding of the effects of these ambiguities.
The position of the oscillation peak moves around in the $E$-$P$ plane
when the value of a parameter is ambiguous and varies within its
allowed region.
The peak position will thus be smeared and consequently the loop of
the peak will appear broadened and blurred.
The value of $\delta$ will be accordingly obscured and the mass
hierarchy will be misidentified within the extended region around the
intersection of the loops.
In addition, the broadening of a narrow loop makes its parallel sides
indistinguishable.
We hence expect that ambiguities of the mass parameters and the mixing
parameters will worsen the degeneracy of the target parameters and
hinder attempts to determine them.
We need extra effort to keep the ambiguities under control, such as
combining with the $\nu_{\mu} \to \nu_{\mu}$ disappearance channel or
employing reactor neutrino experiments.
Detailed analyses of the ambiguities in the framework of the peak
loops are left for future works.

The present analysis based on the peak loops for the oscillation
probability is independent of experimental specifications and gives an
organized understanding of the presence and absence of degeneracy.
A more practical evaluation of quantitative capabilities can be
carried out for specific experiments by applying the expected number of
events to our analysis in place of the oscillation probability.
For that purpose, we need to calculate the expected number of events
by using the neutrino beam flux of the experiment, the neutrino cross
sections, the detector design including a knowledge of systematic
errors and background, and other aspects of the experimental setup.
Drawing the loops traced by the peaks of the spectrum of the event
number, we can study the possibility of the presence of degeneracy,
quantitatively estimate the allowed region in the parameter space, and
select an appropriate baseline length to determine the parameter
values.

\section*{Acknowledgements}
  The authors are grateful to Professor Joe Sato for his encouragement
  during this work.

\appendix

\section{Oscillation Probabilities to Second Order}
\label{app-sec:AKS-approx-deriv}
We calculate the $S$-matrix of neutrino oscillation to second
order in $\Delta_{21} \equiv \delta m^{2}_{21}L/2E$ and
$\Delta_{\textrm{m}} = aL/2E$, and also derive a formula for the
oscillation probability to this order.

The evolution equation of neutrinos
$\nu = (\nu_{\textrm{e}}, \nu_{\mu}, \nu_{\tau})^{\textrm{T}}$ is
given by
\begin{equation}
  i \frac{d \nu(x)}{d x}
  =
  H(x) \nu(x) \, ,
\end{equation}
where
\begin{equation}
\begin{split}
  H(x)
  =
  \frac{1}{2E}
  \Bigl(
  &
    U \mathrm{diag} (0, \delta m^{2}_{21}, \delta m^{2}_{31}) U^{\dagger}
    +
    \mathrm{diag} (a(x), 0, 0)
  \Bigr) \, .
\end{split}
\end{equation}
Here %
$E$ is the neutrino energy, %
$U$ is the unitary mixing matrix of the neutrinos (MNS matrix), %
and $a(x) = 2\sqrt{2} G_{\textrm{F}} n_{\textrm{e}}(x) E$ represents
the matter effect, %
where $G_{\textrm{F}}$ is the Fermi constant %
and $n_{\textrm{e}}(x)$ is the number density of electrons on the
baseline.
The probability for a neutrino $\nu_{\alpha}$ to change into
$\nu_{\beta}$ ($\{\alpha, \beta\} \subset \{\mathrm{e}, \mu, \tau\}$)
is given by
\begin{equation}
  P(\nu_{\alpha} \to \nu_{\beta})
  = \big\lvert S(x)_{\beta \alpha} \big\rvert^{2} \, ,
  \label{eq:P-is-SS}
\end{equation}
where
\begin{equation}
  S(x) =
  \mathrm{T} \exp
  \bigg( -i \int_{0}^{x} d s \, H(s) \biggr) \, .
  \label{eq:Smatrix-def1}
\end{equation}
We assume in this paper that the spatial variation of the electron
density $n_{\textrm{e}}$ is negligible so that
Eq.~(\ref{eq:Smatrix-def1}) simplifies to
\begin{equation}
  S(x) = e^{-i H x} \, .
  \label{eq:Smatrix-def2}
\end{equation}
We split $H$ as $H = H_{0} + H_{1}$, where we have
\begin{align}
  H_{0} & \equiv
  \frac{1}{2E} U \mathrm{diag} (0, 0, \delta m^{2}_{31}) U^{\dagger} \, ,
  \\
  H_{1} & \equiv
  \frac{1}{2E} 
  \Bigl( U \mathrm{diag} (0, \delta m^{2}_{21}, 0) U^{\dagger} +
  \mathrm{diag} (a, 0, 0) \Bigr) \, ,
\end{align}
and treat $H_{1}$ as a perturbation to obtain
\begin{equation}
  S(x) =
  e^{-i H_{0} x} 
  \mathrm{T} \exp \biggl(
    -i \int_{0}^{x} d s \,
    e^{i H_{0} s} H_{1} e^{-i H_{0} s}
  \biggr) \, .
  \label{eq:AKS-Smatrix}
\end{equation}
Expansion of Eq.~(\ref{eq:AKS-Smatrix}) leads to a systematic
approximation of $S$-matrix.\cite{AKS-related} \
An explicit calculation of $S(x)$ up to the second order in $H_{1}$
gives
\begin{equation}
  S(x)_{\beta \alpha} =
  S_{0}(x)_{\beta \alpha} + S_{1}(x)_{\beta \alpha} + S_{2}(x)_{\beta \alpha} \, ,
\end{equation}
where
\begin{equation}
  S_{0}(x)_{\beta \alpha} =
  \delta_{\alpha \beta}
  - U_{\beta 3} U^{\ast}_{\alpha 3}
    \bigl( 1 - e^{-i \Delta_{31}} \bigr) \, ,
  \label{eq:AKS-S0}
\end{equation}
\begin{equation}
\begin{split}
  S_{1}(x)_{\beta \alpha}
  = &
  - i \Delta_{21} U_{\beta 2} U^{\ast}_{\alpha 2}
  \\ &
  - i \Delta_{\textrm{m}}
  \bigg[
      \delta_{\beta \mathrm{e}} \delta_{\alpha \mathrm{e}}
    - (\delta_{\alpha \mathrm{e}} + \delta_{\beta \mathrm{e}})
      U_{\beta 3} U^{\ast}_{\alpha 3}
    + |U_{\mathrm{e} 3}|^{2} U_{\beta 3} U^{\ast}_{\alpha 3}
      \big( 1 + e^{-i \Delta_{31}} \big)
  \bigg]
  \\
  &
  -
  \frac{\Delta_{\textrm{m}}}{\Delta_{31}}
  \big(
  \delta_{\alpha \mathrm{e}}+ \delta_{\beta \mathrm{e}}
  - 2 |U_{\mathrm{e} 3}|^{2}
  \big)
  U_{\beta 3} U^{\ast}_{\alpha 3}
  \big( 1 - e^{-i \Delta_{31}} \big) \, ,
\end{split}
\label{eq:AKS-S1}
\end{equation}
and
\begin{equation}
\begin{split}
  &
  S_{2}(x)_{\beta \alpha}
  =
  - \frac{1}{2}
  \Delta_{21}^{2}
  U_{\beta 2} U^{\ast}_{\alpha 2}
  \\ & \quad
  -
  \Delta_{\textrm{m}} \Delta_{21}
  \bigg[
    \frac{1}{2}
    (\delta_{\alpha \mathrm{e}} + \delta_{\beta \mathrm{e}})
    U_{\beta 2} U^{\ast}_{\alpha 2}
    \\ & \hspace{20mm}
    -
    (
      U_{\mathrm{e} 2} U^{\ast}_{\mathrm{e} 3} U_{\beta 3} U^{\ast}_{\alpha 2}
      +
      U_{\mathrm{e} 3} U^{\ast}_{\mathrm{e} 2} U_{\beta 2} U^{\ast}_{\alpha 3}
    )
    \bigg(
      \frac{1}{2}
      + \frac{i}{\Delta_{31}}
      - \frac{1 - e^{- i \Delta_{31}}}{\Delta_{31}^{2}}
    \bigg)
  \bigg]
  \\ & \quad
  -
  \Delta_{\textrm{m}}^{2}
  \mathbf{\Bigg(}
    \frac{1}{2}
    \delta_{\beta \mathrm{e}} \delta_{\alpha \mathrm{e}}
    -
    ( \delta_{\alpha \mathrm{e}} + \delta_{\beta \mathrm{e}} )
    U_{\beta 3} U^{\ast}_{\alpha 3}
    \bigg(
      \frac{1}{2}
      + \frac{i}{\Delta_{31}}
      - \frac{1 - e^{- i \Delta_{31}}}{\Delta_{31}^{2}}
    \bigg)
    \\ & \hspace*{13mm}
    +
    |U_{\mathrm{e} 3}|^{2}
    \bigg\{
      - \delta_{\beta \mathrm{e}} \delta_{\alpha \mathrm{e}}
      \bigg(
        \frac{1}{2}
        + \frac{i}{\Delta_{31}}
        - \frac{1 - e^{- i \Delta_{31}}}{\Delta_{31}^{2}}
      \bigg)
      \\ & \hspace*{18mm}
      +
      (1 + \delta_{\alpha \mathrm{e}} + \delta_{\beta \mathrm{e}})
      U_{\beta 3} U^{\ast}_{\alpha 3}
      \bigg[
        \frac{1}{2}
        +
        \frac{i
              \big( 2 + e^{-i \Delta_{31}} \big)}
             {\Delta_{31}}
        -
        \frac{3 \big( 1 - e^{- i \Delta_{31}} \big)}
             {\Delta_{31}^{2}}
      \bigg]
    \bigg\}
    \\ & \hspace*{13mm}
    -
    |U_{\mathrm{e} 3}|^{4} U_{\beta 3} U^{\ast}_{\alpha 3}
    \bigg[
      \frac{1 - e^{- i \Delta_{31}}}{2}
      + \frac{3 i \big( 1 + e^{- i \Delta_{31}} \big)}{\Delta_{31}}
      - \frac{6 \big( 1 - e^{- i \Delta_{31}} \big)}{\Delta_{31}^{2}}
    \bigg]
  \mathbf{\Bigg)} \, ,
\end{split}
\label{eq:AKS-S2}
\end{equation}
with $\Delta_{ij} \equiv \delta m^{2}_{ij} L/2E$ and
$\Delta_{\textrm{m}} \equiv aL/2E$.
Substituting Eqs.~(\ref{eq:AKS-S0}), (\ref{eq:AKS-S1}) and (\ref{eq:AKS-S2})
into Eq.~(\ref{eq:P-is-SS}), we obtain the expression for the
oscillation probability up to second order as
\begin{subequations}
\begin{equation}
  P(\nu_{\alpha} \to \nu_{\beta}) =
    p_{0}(\nu_{\alpha} \to \nu_{\beta})
  + p_{1}(\nu_{\alpha} \to \nu_{\beta})
  + p_{2}(\nu_{\alpha} \to \nu_{\beta}) \, ,
\end{equation}
where
\begin{equation}
  p_{0}(\nu_{\alpha} \to \nu_{\beta})
  =
    \delta_{\alpha \beta}
  - 4 |U_{\beta 3}|^{2}
    \big( \delta_{\alpha \beta} - |U_{\alpha 3}|^{2} \big) s^{2} \, ,
\end{equation}
\begin{equation}
\begin{split}
  p_{1}(\nu_{\alpha} \to \nu_{\beta})
  &
  =
  4
  \Delta_{21}
  \Big[
      \mathrm{Re} \, (U^{\ast}_{\beta 3} U_{\beta 2} U_{\alpha 3} U^{\ast}_{\alpha 2})
      s c
    - \mathrm{Im} \, (U^{\ast}_{\beta 3} U_{\beta 2} U_{\alpha 3} U^{\ast}_{\alpha 2})
      s^{2}
  \Big]
  \\ &
  + 
  4
  \Delta_{\textrm{m}}
  |U_{\mathrm{e} 3}|^{2}
  \big(
      \delta_{\alpha \mathrm{e}} \delta_{\beta \mathrm{e}}
    - \delta_{\alpha \mathrm{e}} |U_{\beta 3}|^{2}
    - \delta_{\beta \mathrm{e}} |U_{\alpha 3}|^{2}
  \\ & \hspace{15mm}
    - \delta_{\alpha \beta} |U_{\beta 3}|^{2}
    + 2 |U_{\beta 3}|^{2} |U_{\alpha 3}|^{2}
  \big)
  \bigg( sc - \frac{2}{\Delta_{31}} s^{2} \bigg) \, ,
\end{split}
\end{equation}
and
\begin{equation}
   p_{2}(\nu_{\alpha} \to \nu_{\beta}) 
   =
   \Delta_{21}^{2} \, p_{2\textrm{A}}
   +
   \Delta_{21} \Delta_{\textrm{m}} \, p_{2\textrm{B}}
   +
   \Delta_{\textrm{m}}^{2} \, p_{2\textrm{C}},
\label{eq:P-aks-order2}   
\end{equation}
with
\begin{equation}  
\begin{split}
  p_{2\textrm{A}}
  & =
    - \big( \delta_{\alpha \beta} - |U_{\alpha 2}|^{2} \big) |U_{\beta 2}|^{2}
    + 2
      \, \mathrm{Re} \,
      (U^{\ast}_{\beta 3} U_{\beta 2} U_{\alpha 3} U^{\ast}_{\alpha 2})
      s^{2}
    \\ & \hspace*{5mm}
    + 2
      \, \mathrm{Im} \,
      (U^{\ast}_{\beta 3} U_{\beta 2} U_{\alpha 3} U^{\ast}_{\alpha 2})
      sc,
\end{split}
\end{equation}
\begin{equation}
\begin{split}
 p_{2\textrm{B}} 
 & =
  -
  2
  \big(
  \delta_{\alpha \mathrm{e}} + \delta_{\beta \mathrm{e}} - 2 |U_{\mathrm{e} 3}|^{2}
  \big)
  \, \mathrm{Re} \,
  (U^{\ast}_{\beta 3} U_{\beta 2} U_{\alpha 3} U^{\ast}_{\alpha 2})
  \bigg( 1 - s^{2} - \frac{2}{\Delta_{31}} sc \bigg)
  \\ & \hspace*{5mm}
  +
  2
  \delta_{\alpha \beta}
  \, \mathrm{Re} \,
  ( U^{\ast}_{\beta 3} U_{\beta 2} U_{\mathrm{e} 3} U^{\ast}_{\mathrm{e} 2} )
  \bigg( 1 - \frac{4}{\Delta_{31}^{2}} s^{2} \bigg)
  \\ & \hspace*{5mm}
  -
  2
  \Big[
    |U_{\beta 3}|^{2}
    \, \mathrm{Re} \,
    ( U^{\ast}_{\mathrm{e} 3} U_{\mathrm{e} 2} U_{\alpha 3} U^{\ast}_{\alpha 2} )
    +
    |U_{\alpha 3}|^{2}
    \, \mathrm{Re} \,
    ( U^{\ast}_{\beta 3} U_{\beta 2} U_{\mathrm{e} 3} U^{\ast}_{\mathrm{e} 2} )
  \Big]
  \\ & \hspace*{60mm} \times
  \bigg( s^{2} + \frac{2}{\Delta_{31}} sc - \frac{4}{\Delta_{31}^{2}} s^{2} \bigg)
  \\ & \hspace*{5mm}
  -
  2
  \Big[
    |U_{\beta 3}|^{2}
    \, \mathrm{Im} \,
    ( U^{\ast}_{\mathrm{e} 3} U_{\mathrm{e} 2} U_{\alpha 3} U^{\ast}_{\alpha 2} )
    +
    |U_{\alpha 3}|^{2}
    \, \mathrm{Im} \,
    ( U^{\ast}_{\beta 3} U_{\beta 2} U_{\mathrm{e} 3} U^{\ast}_{\mathrm{e} 2} )
    \\ & \hspace*{15mm}
    - \big(
    \delta_{\alpha \mathrm{e}} + \delta_{\beta \mathrm{e}} - 2 |U_{\mathrm{e} 3}|^{2}
    \big)
  \, \mathrm{Im} \,
  ( U^{\ast}_{\beta 3} U_{\beta 2} U_{\alpha 3} U^{\ast}_{\alpha 2} )
  \Big]
  \bigg( sc - \frac{2}{\Delta_{31}} s^{2} \bigg) \, ,
\end{split}
\end{equation}
and
\begin{equation}
\begin{split}
  & p_{2\textrm{C}}\\
  & =
  -
  |U_{\mathrm{e} 3}|^{2}
  \big(
      \delta_{\alpha \mathrm{e}} \delta_{\beta \mathrm{e}}
    - \delta_{\alpha \mathrm{e}} |U_{\beta 3}|^{2}
    - \delta_{\beta \mathrm{e}} |U_{\alpha 3}|^{2}
    \big)
  \bigg(
  1 - 2s^{2} - \frac{8}{\Delta_{31}} sc + \frac{12}{\Delta_{31}^{2}} s^{2}
  \bigg)
  \\ & \hspace*{5mm}
  -
  \delta_{\alpha \beta}
  |U_{\mathrm{e} 3}|^{2} |U_{\beta 3}|^{2}
  \bigg( 1 + \frac{4}{\Delta_{31}} sc - \frac{12}{\Delta_{31}^{2}} s^{2} \bigg)
  \\ & \hspace*{5mm}
  +
  2
  |U_{\mathrm{e} 3}|^{2} |U_{\beta 3}|^{2} |U_{\alpha 3}|^{2}
  \bigg( s^{2} + \frac{6}{\Delta_{31}} sc - \frac{12}{\Delta_{31}^{2}} s^{2} \bigg)
  \\ & \hspace*{5mm}
  +
  2
  \delta_{\alpha \beta}
  |U_{\mathrm{e} 3}|^{4} |U_{\beta 3}|^{2}
  \bigg( s^{2} + \frac{6}{\Delta_{31}} sc - \frac{12}{\Delta_{31}^{2}} s^{2} \bigg)
  \\ & \hspace*{5mm}
  +
  | U_{\mathrm{e} 3} |^{4}
  \big(
  \delta_{\alpha \mathrm{e}} \delta_{\beta \mathrm{e}}
  - \delta_{\alpha \mathrm{e}} |U_{\beta 3}|^{2}
  - \delta_{\beta \mathrm{e}} |U_{\alpha 3}|^{2}
  \big)
  \bigg(
  4 - 6 s^{2} - \frac{28}{\Delta_{31}} sc + \frac{40}{\Delta_{31}^{2}} s^{2}
  \bigg)
  \\ & \hspace*{5mm}
  +
  4 |U_{\mathrm{e} 3}|^{4} |U_{\beta 3}|^{2} |U_{\alpha 3}|^{2}
  \bigg(
  1 - 2 s^{2} - \frac{10}{\Delta_{31}} sc + \frac{16}{\Delta_{31}^{2}} s^{2}
  \bigg) \, .
\end{split}
\end{equation}
\end{subequations}
Here we have used the abbreviations $s = \sin (\Delta_{31}/2)$ and $c
\equiv \cos (\Delta_{31}/2)$.

The lengthy expression given in Eq.~(\ref{eq:P-aks-order2}) has the
form
\begin{equation}
\begin{split}
  P(\nu_{\alpha} \to \nu_{\beta})
  = &
    C_{1} \sin^{2} \frac{\Delta_{31}}{2}
  + C_{2} \sin \frac{\Delta_{31}}{2} \cos \frac{\Delta_{31}}{2}
  + C_{3} \, ,
\end{split}
\label{eq:P-in-a-nutshell-1}
\end{equation}
where the coefficients $C_{1}$, $C_{2}$, and $C_{3}$ are written in
terms of the elements of the mixing matrix, the mass parameters, and
$\Delta_{\textrm{m}}$.
The coefficient $C_{1}$ is $O(1)$ while $C_{2}$ is first order or
higher, so that Eq.~(\ref{eq:P-in-a-nutshell-1}) reduces to a
two-generation formula in the zeroth-order approximation.
We factor out $\sin (\Delta_{31}/2)$ from the sum of sinusoidal
functions and transform the sum of the sine and cosine terms into a
single sine term as
\begin{subequations}
\begin{equation}
\begin{split}
  &
  P(\nu_{\alpha} \to \nu_{\beta})
  =
  \sqrt{C_{1}^{2} + C_{2}^{2}} \sin \frac{\Delta_{31}}{2}
  \sin \biggl( \frac{\Delta_{31}}{2} + C_{4} \biggr)
  + C_{3} \, ,
\end{split}
\label{eq:P-in-a-nutshell-2}
\end{equation}
\begin{equation}
 C_{4} = \arctan \frac{C_{2}}{C_{1}},
\end{equation}
\end{subequations}
where we take $0 \leq C_{4} \leq \pi$ for $C_{2} \geq 0$ and
$\pi < C_{4} < 2\pi$ for $C_{2} < 0$. 
We rewrite Eq.~(\ref{eq:P-in-a-nutshell-2}) as
\begin{equation}
\begin{split}
  P(\nu_{\alpha} \to \nu_{\beta})
  & =
    \sqrt{C_{1}^{2} + C_{2}^{2}}
    \sin^{2} \biggl( \frac{\Delta_{31}}{2} + \frac{C_{4}}{2} \biggr)
  - \sqrt{C_{1}^{2} + C_{2}^{2}} \sin^{2} \frac{C_{4}}{2}
  + C_{3} \, ,
\end{split}
\label{eq:P-in-a-nutshell-3}
\end{equation}
which reproduces the form of Eq.~(\ref{eq:Pm2e-aks-order2-body}) with
\begin{equation}
  A \equiv \frac{1}{4l} \sqrt{C_{1}^{2} + C_{2}^{2}} \, , \;
  \Theta \equiv \frac{\Delta_{31}}{2} + \frac{C_{4}}{2} \, , \;
  B \equiv
  \frac{1}{4l}
  \biggl(
    C_{3} - \sqrt{C_{1}^{2} + C_{2}^{2}} \sin^{2} \frac{C_{4}}{2}
  \biggr) \, .
\label{eq:AThetaB-in-a-nutshell}
\end{equation}
The expressions of Eqs.~(\ref{eq:Pm2e-aks-order2-A}),
(\ref{eq:Pm2e-aks-order2-Theta}), and (\ref{eq:Pm2e-aks-order2-B}) are
obtained from Eq.~(\ref{eq:AThetaB-in-a-nutshell}) by setting $\alpha
= \mu$ and $\beta = \mathrm{e}$ in $C_{1}$, $C_{2}$, $C_{3}$, and
$C_{4}$ and taking the terms up to the appropriate order.

\section{Validity of the Approximation}
\label{app-sec:AKS-approx-validity}
We made use of approximate formulae for the oscillation probability
in \S\ref{sec:Analytic-expressions}.
We elucidate the requirements for the these formulae to be valid.

The approximation we developed in Appendix \ref{app-sec:AKS-approx-deriv}
requires
\begin{equation}
  \delta m^{2}_{31} \gg a, \delta m^{2}_{21} \, ,
\end{equation}
\begin{equation}
\begin{split}
  \Delta_{\textrm{m}}
  \equiv \frac{aL}{2E}
  = 5.0 \times 10^{-1}
      \biggl( \frac{L}{1000 \, [\mathrm{km}]} \biggr)
      \biggl( \frac{\rho}{2.6 \, [\mathrm{g \, cm^{-3}}]} \biggr)
  \ll 1 \, ,
\end{split}
\end{equation}
and
\begin{equation}
\begin{split}
  \Delta_{21}(E)
  & \equiv
  \frac{\delta m^{2}_{21} L}{2E}
  \\ &
  = 2.1 \times 10^{-1}
    \biggl( \frac{\delta m^{2}_{21}}{8.2 \cdot 10^{-5} [\mathrm{eV^{2}}]} \biggr)
    \biggl( \frac{L}{1000 \, [\mathrm{km}]} \biggr)
    \biggl( \frac{E}{[\mathrm{GeV}]} \biggr)^{-1}
  \ll  1.        
\end{split}
\end{equation}
\label{eq:AKS-validity-cond}
Since neutrino energy we consider in this paper is
\begin{math}
  E
  \approx E_{\textrm{peak},n}
  \approx |\delta m^{2}_{31}|L/[2(2n + 1)\pi],
\end{math}
the conditions given in Eq.~(\ref{eq:AKS-validity-cond}) are satisfied
at least marginally for $L \sim (300 \,\textrm{--}\, 1500)
\mathrm{km}$.

We additionally simplified the expression by dropping higher-order
terms of $\sin \theta_{13}$ taking account of its smallness as
mentioned in \S\ref{sec:Analytic-expressions}.
Taking account of the relation $\Delta_{21} < \Delta_{\textrm{m}}$, we
dropped terms of $O(\sin^{2} \theta_{13})$ in the coefficients of
$\Delta_{21}^{2}$ and of $\Delta_{21}\Delta_{\textrm{m}}$, as well as
terms of $O(\sin^{3} \theta_{13})$ in the coefficients of
$\Delta_{\textrm{m}}^{2}$.
This simplification is valid when the omitted terms are much smaller
than the leading term which is of $O(\sin^{2} \theta_{13})$, or
explicitly when the following conditions are satisfied:
\begin{equation}
\begin{split}
  &
  \frac{ \Delta_{21}(E_{\textrm{peak},n})^{2} \sin^{2} \theta_{13} }
       { \sin^{2} \theta_{13} }
  = 1.1 \times 10^{-2} \cdot (2n + 1)^{2} 
    \frac{
      \biggl(
        \dfrac{ \delta m^{2}_{21} }{8.2 \cdot 10^{-5} \, [\mathrm{eV}^{2}] }
      \biggr)^{2}
    }
    {
      \biggl(
        \dfrac{|\delta m^{2}_{31}|}{2.5 \cdot 10^{-3} \, [\mathrm{eV}^{2}] } 
      \biggr)^{2}
    }
  \ll 1 \, ,
\end{split}
\end{equation}
\begin{equation}
\begin{split}
  \frac{ \Delta_{21}(E_{\textrm{peak},n}) \Delta_{\textrm{m}} \sin^{2} \theta_{13} }
       { \sin^{2} \theta_{13} }
  &
  = 5.2 \times 10^{-2} \cdot (2n + 1)
      \frac{
        \biggl(
          \dfrac{ \delta m^{2}_{21} }{8.2 \cdot 10^{-5} \, [\mathrm{eV}^{2}] }
        \biggr)
      }
      {
        \biggl(
          \dfrac{|\delta m^{2}_{31}|}{2.5 \cdot 10^{-3} \, [\mathrm{eV}^{2}] } 
        \biggr)
      }
  \\ & \hspace*{1.5cm} \times
      \biggl( \frac{L}{1000 \, [\mathrm{km}]} \biggr)
      \biggl( \frac{\rho}{2.6 \,[\mathrm{g \, cm^{-3}}]} \biggr)        
  \ll 1 \, ,
\end{split}
\end{equation}
and
\begin{equation}
\begin{split}
  \frac{ \Delta_{\textrm{m}}^{2} \sin^{3} \theta_{13} }{ \sin^{2} \theta_{13} }
  =
  2.5 \times 10^{-1} \cdot \sin \theta_{13}
  \biggl( \frac{L}{1000 \, [\mathrm{km}]} \biggr)^{2}
  \biggl( \frac{\rho}{2.6 \, [\mathrm{g \, cm^{-3}}]} \biggr)^{2}
  \ll 1. 
\end{split}
\end{equation}
These inequalities are satisfied for allowed values of $\sin^{2}
2\theta_{13} < 0.19$.\cite{Yao:2006px} \

\end{document}